\documentclass[aps,nofootinbib,twocolumn,notitlepage,floatfix,superscriptaddress,secnumarabic]{revtex4-1}
\usepackage{amsmath}
\usepackage{indentfirst}
\usepackage{graphicx,color}
\usepackage{multirow}
\usepackage{times}
\usepackage{bm}
\usepackage{soul}
\usepackage{tabularx}
\usepackage{epstopdf}
\usepackage{enumerate}
\usepackage[normalem]{ulem}
\usepackage{cancel}
\usepackage{color}
\usepackage{float}
 \usepackage{mathrsfs}
\usepackage{comment}
\usepackage{xcolor}
\usepackage{marginnote}
\usepackage{subfigure}

\newcommand{\stkout}[1]{\ifmmode\text{\sout{\ensuremath{#1}}}\else\sout{#1}\fi}

\usepackage[colorlinks, linkcolor=red,
            anchorcolor=blue, citecolor=green]{hyperref}
\usepackage[T1]{fontenc}

\begin{document}

\title{
Jet tagging algorithm of graph network with HaarPooling message passing}

\author{Fei Ma}
%\email{mafei@mails.ccnu.edu.cn}
\affiliation{Key Laboratory of Quark and Lepton Physics (MOE) and Institute of Particle Physics, Central China Normal University, WuHan, 430079, China}

\author{Feiyi Liu}
\email{fyliu@mails.ccnu.edu.cn}
\affiliation{Key Laboratory of Quark and Lepton Physics (MOE) and Institute of Particle Physics, Central China Normal University, WuHan, 430079, China}
\affiliation{Institute for Physics, E{\"o}tv{\"o}s Lor\'and University\\1/A P\'azm\'any P. S\'et\'any, H-1117, Budapest, Hungary}

\author{Wei Li}
\email{liw@mail.ccnu.edu.cn}
\affiliation{Key Laboratory of Quark and Lepton Physics (MOE) and Institute of Particle Physics, Central China Normal University, WuHan, 430079, China}
\affiliation{Max-Planck-Institute  for  Mathematics  in  the  Sciences,  04103  Leipzig,  Germany}

\date{\today}

\begin{abstract}
Recently methods of graph neural networks (GNNs) have been applied to solving the problems in high energy physics (HEP) and have shown its great potential for quark-gluon tagging with graph representation of jet events.  In this paper, we introduce an approach of GNNs combined with a HaarPooling operation to analyze the events, called HaarPooling Message Passing neural network (HMPNet). In HMPNet,  HaarPooling not only extracts the features of graph, but embeds additional information obtained by clustering of k-means of different particle features. We construct Haarpooling from five different features: absolute energy $\log E$, transverse momentum $\log p_T$, relative coordinates $(\Delta\eta,\Delta\phi)$, the mixed ones $(\log E, \log p_T)$ and $(\log E, \log p_T, \Delta\eta,\Delta\phi)$. The results show that an appropriate selection of information for HaarPooling enhances the accuracy of quark-gluon tagging, as adding extra information of $\log P_T$ to the HMPNet outperforms all the others, whereas adding relative coordinates information $(\Delta\eta,\Delta\phi)$ is not very effective. This implies that by adding effective particle features from HaarPooling can achieve much better results than solely pure message passing neutral network (MPNN) can do, which demonstrates significant improvement of feature extraction via the pooling process. Finally we compare the HMPNet study, ordering by $p_T$, with other studies and prove that the HMPNet is also a good choice of GNN algorithms for jet tagging.

\hspace{0.5cm}

\noindent DOI:10.1103/PhysRevD.108.072007

\end{abstract}
\maketitle

\section{Introduction}\label{intro}
\setlength{\parindent}{2em}  As an event in high energy collisions, a jet refers to a collimated spray of hadrons observed by detectors as a signature of quarks and gluons. In Large Hadron Collider (LHC), jets with dynamic information combined from different detector components are experimentally reconstructed by particle flow algorithms~\cite{beaudette2014cms,Sirunyan_2017,Aaboud:2017tm}.
One of the prime study on jet is to specify the origin of a jet from a type of elementary particle, called jet tagging.
Since the character of the source particles can be surmised from the properties of jets, for example, jets initiated by gluons generally have more extensive energy spread than by quarks. The information of these initial elementary particles could facilitate key tasks in high-energy physics (HEP) experiments, such as searching for new particles and estimating the Standard Model processes.
 
In past decades, researches on jet tagging via QCD theory have never stopped and continuously improved for quark and gluon jets~\cite{PhysRevLett.107.172001,Larkoski:2014vc, Bhattacherjee:2015ta,PhysRevD.95.034001,Gras:2017ul}, top jets~\cite{PhysRevLett.101.142001,PhysRevD.85.034029,PhysRevD.87.054012,PhysRevD.89.074047,Kasieczka:2015tn,Thaler:2012tb} and jets from bottom quarks~\cite{Gandara:2009vr,ATLAS:2016wv,Aad:2019we}. Recently, methods of deep learning (DL) have been applied to studying jet classification, by constructing a representation of event paired with a corresponding analysis method, such as particle calorimeter images with convolutional neural networks (CNNs) ~\cite{Oliveira:2016vm,Komiske:2017wc,Macaluso:2018to,Kasieczka:2017vt,Schwartzman:2016tu},  particle lists with recurrent neural networks (RNNs) \cite{Louppe:2019vz,egan2017long,Fraser:2018uo, Cheng:2018tr}, and collections of ordered inputs with dense neural networks (DNNs) \cite{PhysRevD.93.094034,PhysRevD.95.014018,PhysRevD.94.112002}. Moreover, energy flow networks (EFNs) treat jet tagging model under the framework of deep sets, which respect infrared and collinear safety by construction~\cite{Komiske:2019um,PhysRevD.103.074022}.
Interaction networks (INs) also have great potential in identifying all-hadronic decays of high-momentum heavy particles~\cite{Moreno:2020up}.
Compared to previous traditional approaches, methods of DL not only could better handle large amount of sophisticated data generated by modern detectors, but also are powerful in analyzing complex internal relations from limited input, leading to great advantages in dealing with jet tagging.

Previous researches have shown that graph neural networks (GNNs) can well handle collision events~\cite{2018Novel,Arjona-Martinez:2019vx,Qasim:2019ux}. For jet tagging, an event usually contains the information of a set of particles with certain kinematic features. As a sensitive probe for classification, the geometrical relationship between these particles can be represented by a geometrical pattern of multiple entities, i.e., the structure of a graph. This graph representation of jets is very flexible as input to DL, which has clear information of particles and does not require additional sorting or information. In Refs.~\cite{Abdughani:2019uk,Ren:2020tu,henrion2017neural}  the graph representation has been applied to jet classification of high-momentum heavy particles via message passing neutral network (MPNN), an algorithm of GNNs. A similar representation called ``particle cloud'' treats a jet as an unordered set of particles, paired with dynamic graph convolutional neural network (DGCNN) as ParticleNet (PN)~\cite{qu2020jet}. 
Methods of autoencoder based on GNNs are also used to distinguish QCD jets and non-QCD jets~\cite{Atkinson:2021tp,10.3389/frai.2022.943135}. As a framework of GNNs, LundNet~\cite{Dreyer:2021uo,Dreyer:2022ww,Dreyer:2022wk} has been proposed for jet tagging in the Lund plane~\cite{Dreyer:2018ut}, by transforming the Lund tree into a graph. And LorentzNet, a symmetry-preserving model of GNNs, describes the particle cloud representation of a jet by the neural network architecture under Lorentz-equivariant~\cite{Li:2022xfc,Gong:2022wk}.
These successful attempts inspire us to deal with jet tagging problems via graph representations and GNNs.
%~\cite{henrion2017neural,IEEE8614089}~\cite{Wang3326362}

Graph pooling is a technique used to reduce the dimension and extract the features of graphs, which usually appear with the convolutional layers~\cite{NIPS2015_f9be311e}. The most widely used methods are graph clustering algorithms \cite{Kushnir:2006un,ijcai2018p490, NIPS2016_04df4d43, NEURIPS2018_e77dbaf6}, as well as some other ones which have been lately studied \cite{cangea2018towards,noutahi2019t,ma2019graph,Grattarola:2022wv}. 
HaarPooling is a graph pooling operation  to compress and filter graph features~\cite{wang2019haarpooling}, based on compressive Haar transforms. One of its important characteristics is, the basis for forming a Haar matrix is computed by a clustering step from the input graph, which means additional input-related information can be passed to the ML process via the Haar matrix. For quark-gluon tagging using GNNs, 
HaarPooling makes it possible to embed extra particle features to filter and enhance the message passing.

In our work, we combine HaarPooling with MPNN to build a new network structure, called HaarPooling Message Passing neural network (HMPNet). On one hand, jet events are transformed into a graph representation as input for GNN, and the tagging can be achieved by training with the process of message passing and self-updating~\cite{Ren:2020tu,qu2020jet}. On the other hand, in the updating process of the algorithm, the additional particle feature is embedded through the compressive Haar basis matrix of pooling, which makes the extraction and classification of features more relevant to the input. This means the pooling for compression also becomes an operation for adding fine information of input. For test, we implement the HMPNet to the quark-gluon tagging of the process $pp \to Z / \gamma^{\ast} + j + X \to \mu^{+} \mu^{-} + j + X$,  and use different particle features such as absolute energy $\log E$, transverse momentum $\log p_{T}$, the relative coordinates $(\Delta\eta,\Delta\phi)$, the mixed ones $(\log E, \log p_T)$ and $(\log E, \log p_T, \Delta\eta,\Delta\phi)$ to generate the Haar matrix by clustering the input, respectively. We analyse the influences of different particle features, and compare the results of  $\log p_T$ with the counterparts of other algorithms, which shows a remarkable improvement of performance.

The main structure of this paper is as follows. In Section \ref{GrapRep}, the graph representation of jets will be given. Section \ref{MPNN} gives the method of MPNN. Section \ref{Haar} includes the conceptions of graph pooling and Haar matrix. In Section \ref{Cluster}, the method of embedding particle features to Haar matrix is illustrated. In Section \ref{NetFlow}, we explain the detailed process of HMPNet. In Section \ref{Data}, the input data and settings of HMPNet are listed. Section \ref{Results} shows our major findings. Section \ref{Conclusion} is the conclusion of this work.

\section{Methodology}\label{meth}

\subsection{Graph representation of jets}\label{GrapRep}

In the language of GNNs, an undirected graph $\mathcal{G}=\{\mathcal{V}, \mathcal{E}, \mathcal{X}, \mathcal{W}\}$ is defined with 
 nodes (vertices) $\mathcal{V}$, edges $\mathcal{E}$, weights of nodes $\mathcal{X}$ and of edges $\mathcal{W}$. Each node $v_i\in\mathcal{V}$ has its feature vector $x_i\in\mathcal{X}$, and for the edge weight $\mathcal{W}$, it is always given in the form of an weight matrix $d_{ij}$ in which the element is given for the edge between $i$-th and $j$-th nodes in the graph. And the number of nodes is defined as $N=|\mathcal{V}|$.

Usually, the information of a jet reconstructed from detectors in high-energy collision includes: the three Cartesian coordinates of the momentum $( p_x, p_y,p_z )$, the absolute energy $E$, the pseudorapidity $\eta$, the azimuthal angle $\phi$, the transverse momentum $p_T$ and so forth. For the feature vectors $\boldsymbol{x_{i}}$, we use 10 variables of jet information as components of $\boldsymbol{x_{i}}$ similar to Ref.~\cite{qu2020jet}, as shown in Table.~\ref{Input_info}. The dynamic information of objects includes $\log p_T$, $\log E$, the relative energy $\log \frac{E}{E(jet)}$ and the relative transverse $\log \frac{p_T}{p_T(jet)}$. In addition, $q$ denotes the electric charge of object and the rest four features are particles identity (PID) information. The dimension of $\boldsymbol{x_{i}}$ is $N\times d_x$, where $d_x = 10$ is the dimension of the feature space.

\begin{figure*} %nput_struc.pdf
\includegraphics[scale=1.2]{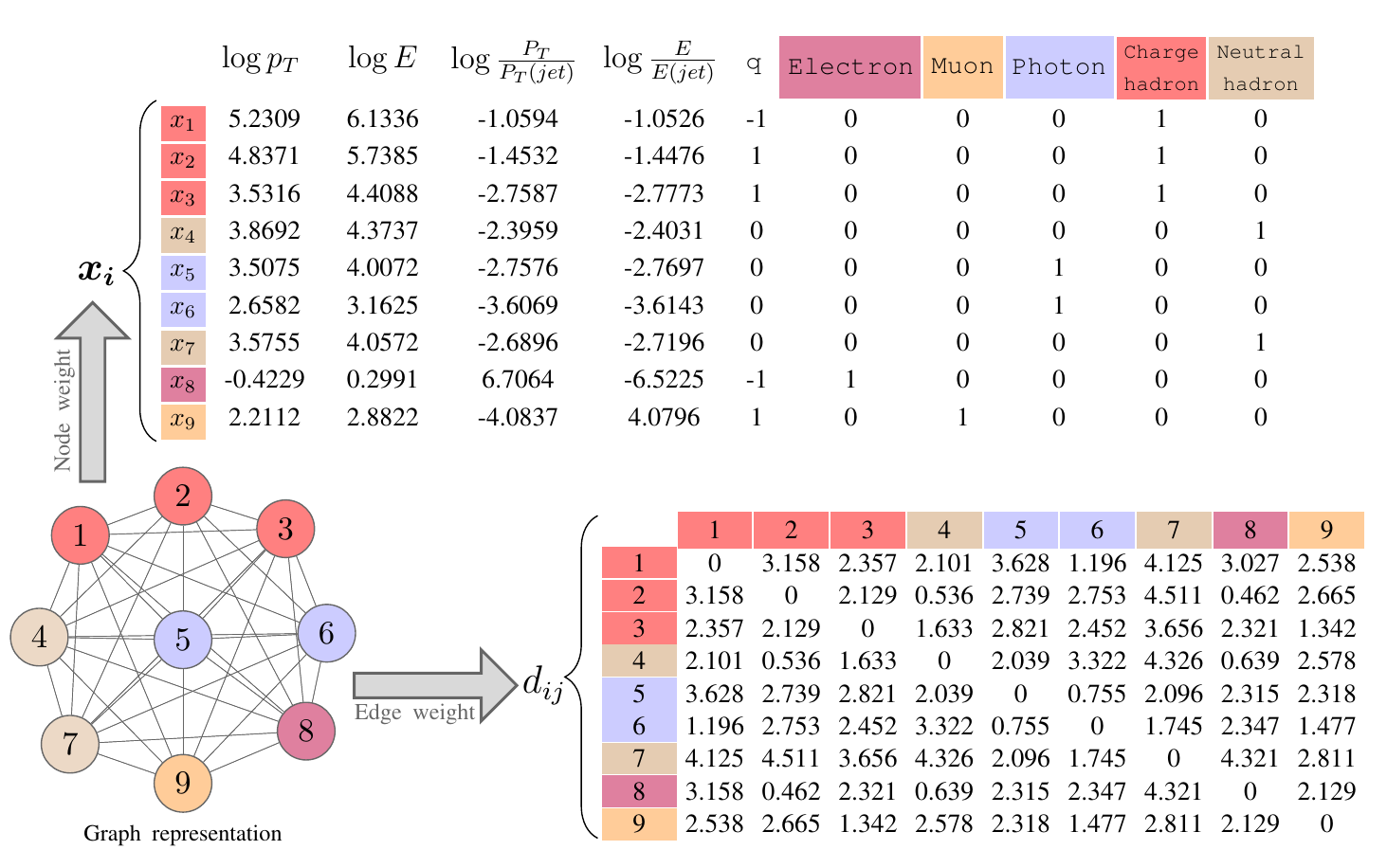}\caption{An event graph with node and edge weights for a specific simulated event of the process $pp \to Z / \gamma^{\ast} + j + X \to \mu^{+} \mu^{-} + j + X $.}\label{fig:input_struc}
\end{figure*}

For graph representation, we also need to identify a parameter as the edge weight $d_{ij}$. From the point view of jet axis, the relative distance $\Delta R = \sqrt{\Delta \eta^2_{ij} + \Delta \phi^2_{ij}}$ from the jet center is a suitable choice, where the relative coordinates $\Delta \eta_{ij} = \eta_i - \eta_j$ and $\Delta \phi_{ij} = \phi_i - \phi_j$ denote the angle difference between the $i$-th with $j$-th  particle in jet axis. 
By the definition of $\Delta R$, the edge weight is given by,
\begin{equation}
d_{ij} = \sqrt{\Delta \eta^{2}_{ij} + \Delta \phi^{2}_{ij}}.
\end{equation}
As an illustration, we show the graph events of the process $pp \to Z / \gamma^{\ast} + j + X \to \mu^{+} \mu^{-} + j + X $ by Monte Carlo simulations in Fig. \ref{fig:input_struc}. As a graph representation with $N$ nodes, each component of $\boldsymbol{x_{i}}$ is a vector of $d_x$ elements of jet information, with $N=9$ and $d_x = 10$. So $d_{ij}$ is an $N\times N$-dimensional symmetric, matrix with all the diagonal elements being 0. Since $\phi$ is not encoded in the node features, the graph representation is invariant under rotation in $\phi$.

\begin{table*}[htbp]
\centering
\caption{Input variables used in the quark-gluon tagging task with PID information.}
\label{tab:input-features}
\resizebox*{\linewidth}{!}{
\begin{ruledtabular}
\begin{tabular}{ccc}
Feature of graph representation       & Variable                                  & Definition                                                                       \\ \cline{1-3} 
%\cline{1-3} 
\multirow{4}{*}{}         & $\log p_T$                                & logarithm of the particle's $p_{T}$                                              \\
                          & $\log E$                                  & logarithm of the particle's energy                                               \\
                          & $\log \frac{p_T}{p_T(\text{jet})}$        & logarithm of the particle's $p_{T}$ relative to the jet $p_{T}$                  \\
                          & $\log \frac{E}{E(\text{jet})}$            & logarithm of the particle's energy relative to the jet energy                    \\
\boldsymbol{$x_i$}    & \texttt{q}                                       & electric charge of the particle                                                  \\
\multirow{5}{*}{} & \texttt{isElectron}      & $1$ if the particle is an electron else $0$                                      \\
                          & \texttt{isMuon}          & $1$ if the particle is a muon else $0$                                           \\
                          & \texttt{isChargedHadron} & $1$ if the particle is a charged hadron else $0$                                 \\
                          & \texttt{isNeutralHadron} & $1$ if the particle is a neutral hadron else $0$                                 \\
                          & \texttt{isPhoton}        & $1$ if the particle is a photon else $0$                                \\ \cline{1-3} 
\multicolumn{1}{c}{\multirow{2}{*}{$d_{ij}$}} & $\Delta \eta_{ij}$   & difference in pseudorapidity between the $i$-th and $j$-th particle in jet axis  \\
                          & $\Delta \phi_{ij}$                        & difference in azimuthal angle between the $i$-th and $j$-th particle in jet axis                                    
\end{tabular}
\end{ruledtabular}
}
\label{Input_info}
\end{table*}
 
\subsection{MPNN algorithm}\label{MPNN}
The flexible and complete feature of graph makes it a natural and promising representation of jets; on the other hand, to choose a paired algorithm of GNNs also requires careful thought. Message Passing Neural Networks(MPNN) is introduced as a powerful and efficient supervised algorithm of GNNs which can learn geometric representations as well, especially the edge features $d_{ij}$~\cite{henrion2017neural,pmlr-v70-gilmer17a}. By finding the optimized  parameters in the nonlinear network model via training, one can obtain the classification as output of MPNN, from the input graph representation of jets.

To start the process of MPNN, the feature vectors $\boldsymbol{x}_{i} \in  R^{N\times d_{x}}$ are embedded into a matrix consisting of higher dimensional state vectors $\boldsymbol{s}_{i}^{(0)} \in R^{N\times d_{s}}$ with $d_{s} >d_{x}$, by an embedding function $f_{e}$:
\begin{equation}
\boldsymbol{s}_{i}^{(0)} = f_{e}(\boldsymbol{x}_{i}).
\label{s_i}
\end{equation}
Here $\boldsymbol{s}_{i}^{(0)}$ is only related to $\boldsymbol{x}_{i}$ without any information of the graph structure. 
To encode the whole event graph into each node state vector, message vector $\boldsymbol{m}_{i}^{(t)}$ is introduced to pass the message of  $\boldsymbol{s}_{i}^{(t-1)}$ and edge weight $d_{ij}$ via the message passing function $f_m$ in the $t$-th iteration as
\begin{equation}
\boldsymbol{m}_{i}^{(t)} = \sum_{j \neq i} \boldsymbol{m}_{i \gets j}^{(t)} = \sum_{j \neq i} f_{m}^{(t)}(\boldsymbol{s}_{j}^{(t-1)},\boldsymbol{d}_{ij}),
\label{m_i}
\end{equation}	
and update its state vector
\begin{equation}
\boldsymbol{s}_{i}^{(t)} = f_{u}^{(t)}(\boldsymbol{s}_{i}^{(t-1)},\boldsymbol{m}_{i}^{(t)}),
\label{s_vector}
\end{equation}
where $f_{u}^{(t)}$ is the update function. This is how a node $i$ collects the messages sent from other nodes in the $t$-th iteration in the message passing layer. 

By the repetition of the message passing procedure, the information of feature from each node and edge continuously passes to the other ones, until each node state contains the information of all other nodes and relations in the entire graph after $T$ iterations. At this time, they can be regarded as the event features automatically extracted from the input event graph. Next, each node votes a number as the likeness of the event to be signal-like, based on its own state vector. Meanwhile, the signal-like probability $y$ is calculated by the voting function averaged over the number of nodes $N=|\mathcal{V}|$ as
\begin{equation}
y = \frac{1}{N}\sum_i f_{v}(\boldsymbol{s}_i^{(T)}).
\label{y_score}	
\end{equation}
The whole ML process directly extracts features to discrimination scores $y$ from event graphs. However, the MPNN only performs feature extraction through message passing, and does not have an extra process of feature compression and filtering, which affects the quality of the extracted features for the classification of output. Usually, a pooling layer is added to GNNs to reduce the dimension of the previous process. Here we would like to combine a HaarPooling operation with the MPNN as a graph pooling process, not only to improve the capability of feature extraction, but to embed additional raw information on particle features for enhancing the message passing and updating. The additional information of particles added during ML process can provide richer content for message passing and guide feature extraction for classification, according to the filtering and compression of the pooling operator. Since the particle features behind jet events have very complex correlations for classification, it is necessary to enhance feature passing, filtering and updating in the process of GNNs to deal with this kind of problems.

 \subsection{Haar graph pooling}\label{Haar}
 
 Graph pooling methods are aiming at graph structure reduction, which eases the diffusion of context between nodes further in the graph. It is also known as graph coarsening, which starts with a coarser version of the graph~\cite{Bacciu:2021wg}. A coarse-grained graph $\mathcal{G}_c=\{\mathcal{V}_c, \mathcal{E}_c, \mathcal{X}_c, \mathcal{W}_c\}$ of $\mathcal{G}=\{\mathcal{V}, \mathcal{E}, \mathcal{X}, \mathcal{W}\}$ means that, each node $v^{(c)}\in\mathcal{V}_c$ is a cluster of $\mathcal{G}$: $v^{(c)}=\{v\in V|v \text{ has parent } v^{(c)}\}$ if $|V_c|\leq|V|$. Each node of $\mathcal{G}_c$ is called a cluster of $\mathcal{G}$. Then a graph coarsening chain  $\mathcal{G}_{0\to J} := (\mathcal{G}_{0},\mathcal{G}_{1},\cdots,\mathcal{G}_{J})$ can be built from the original graph $\mathcal{G}_0$ to the $J$-th coarsened graph $\mathcal{G}_J$, where $J$ is a positive integer~\cite{auer2012graph}. 
 
Ordering $\mathcal{V}_c$ by node weights or other related quantities as $\mathcal{V}_c=\{v_1^{(c)},\cdots,v_{N_c}^{(c)}\}$ for $N_c=|\mathcal{V}_c|$, the vectors $\boldsymbol{\phi}_l^{(c)}\in R^{N_c}$ on $\mathcal{G}_{c}$ can be defined as
\begin{equation}
\boldsymbol{\phi}_1^{(c)}(v^{(c)})=\frac{1}{\sqrt{N_c}}, \quad v^{(c)}\in \mathcal{V}_c,
\label{Haar1}
\end{equation}
and for $l=2,\cdots,N_c$,
\begin{equation}
\boldsymbol{\phi}_l^{(c)}(v^{(c)})=\sqrt{\frac{N_c-l+1}{N_c-l+2}}\left(\chi_{l-1}^{(c)}-\frac{\sum_{j=l}^{N_c}\chi_j^{(c)}}{N_c-l+1}\right),
\end{equation}
where the indicator function for the $j$-th vertex $v^{(j)}\in V_j$ on $\mathcal{G}_{j-1}$ is given by
\begin{equation}
\chi_j^{(c)}(v^{(c)}) =
  \begin{cases}
1, & v^{(c)}=v_j^{(c)}, \\
 0, & v^{(c)}\in  \mathcal{V}_c \setminus \{v_j^{(c)}\}.
  \end{cases}       
\end{equation}
Then, an orthonormal basis can be formed by the $\{ \boldsymbol{\phi}_l^{(c)}\}_{l=1}^{N_c}$ as each $v\in \mathcal{V}$ belongs to an exact cluster $v_c\in\mathcal{V}_c$~\cite{wang2019haarpooling}. So each vector $\boldsymbol{\phi}_l^{(c)}$ on $\mathcal{G}_c$ can be expressed as a vector on $\mathcal{G}$ by
\begin{equation}
\boldsymbol{\phi}_{l,1}(v) = \frac{\boldsymbol{\phi}_l^{(c)}(v^{(c)})}{\sqrt{|v^{(c)}|}}, \quad v\in v^{(c)} \text{ and } l=1,\cdots,N_c.
\end{equation}
As the cluster size $|v^{(c)}|=m_l$ is also the number of nodes in $\mathcal{G}$ having common parents $v^{(c)}$. By ordering $v_l^{(c)}=\{v_{l,1},\cdots,v_{l,m_l}\}\subseteq \mathcal{V}$, the orthonormal basis for $m=2,\cdots,m_l$ is defined as
\begin{equation}
\boldsymbol{\phi}_{l,m} = \frac{m_l-m+1}{m_l-m+2}\left(\chi_{l,m-1}-\frac{\sum_{j=m}^{m_l}\chi_{l,j}}{m_l-m+1}\right),
\label{Haar4}
\end{equation}
where
\begin{equation}
\chi_{l,j}(v) =
  \begin{cases}
1, & v=v_{l,j}, \\
 0, & v\in  \mathcal{V}\setminus \{v_{l,j}\},
  \end{cases}
  \quad j = 1,\cdots,m_l.       
\end{equation}
With this orthonormal basis $ \boldsymbol{\phi}_{l,m}$ for $l=1,\cdots,N_c$ and $m=1,\cdots,m_l$, the Haar basis for the $j$-th layer can be defined as  $\{ \boldsymbol{\phi}_l^{(j)}\}_{l=1}^{N_j}$ in a chain $\mathcal{G}_{0\to J}$ for $j=0,\cdots,J-1$, by repeating the generation steps from Eq.~(\ref{Haar1}) to~(\ref{Haar4}) for $0\to j$-th layers. And the compressed Haar basis is $\{ \boldsymbol{\phi}_l^{(j)}\}_{l=1}^{N_{j+1}}$ for $N_{j+1}\leq N_j$.
 
For the pooling process, a dimensionality reduction from $\mathcal{G}_j\in R^{N_j}$ to $\mathcal{G}_{j+1}\in R^{N_{j+1}}$ for $N_{j+1}<N_j$ requires a mapping from elements of an $N_j$ size vector to $N_{j+1}$, which can be achieved by a transformation matrix. For this purpose, here we introduce the Haar basis matrix for the $j$-th layer $\Phi^{(j)}_{N_{j}\times N_{j}} = \{\boldsymbol{\phi}_{1}^{(j)},\cdots,\boldsymbol{\phi}_{N_{j}}^{(j)}\} \in R^{N_{j}\times N_{j}}$ based on $\{ \boldsymbol{\phi}_l^{(j)}\}_{l=1}^{N_{j}}$,  which also can be shortly written as $\tilde{\Phi}_{j}$, and the compressive Haar basis matrix $\Phi^{(j)}_{N_{j+1}\times N_{j}}$ as $\Phi_{j}$ for $N_{j+1}<N_j$~\cite{wang2019haarpooling}. This transformation process is called ``HaarPooling'' for a GNN with $K$ pooling layers as
 \begin{equation}
 X_{\text{out}} =\Phi_{j}^T X_{\text{in}}, \quad j=0,1,\cdots,K-1,
\end{equation}  
 where the input feature array $X_{\text{in}}\in R^{N_{j+1}\times d}$ and the output $X_{\text{out}}\in R^{N_{j}\times d}$ for $N_{j+1}<N_j$. For the last graph of chain, $\mathcal{G}_J$, $N_K = |\mathcal{V}_J|=1$. 
 
 By the definition, we know that the nodes $\{v_1^{(c)},\cdots,v_{N_c}^{(c)}\}$ forming the Haar basis  $\{\boldsymbol{\phi}_{1}^{(c)},\cdots,\boldsymbol{\phi}_{N_{c}}^{(c)}\}$ can be determined by the ordering of node weights. It means that the values of the Haar matrix are different, depending on the sorting methods. This allows us to reorder and classify nodes according to different values of weights, and pass the information of distribution or labels of clusters to enhance feature extraction during GNN process through the Haar matrix. In our study, the jet events are clustered ones according to different information of particles passing to pooling by $\Phi_j$, which will be explained as follows.
 
 \subsection{The cluster information of particles}\label{Cluster}
 
 According to the information of jet events for graph representation in Table~\ref{Input_info}, we choose the features  commonly used in HEP for clustering:  absolute energy $\log E$, transverse momentum $\log p_{T}$, and relative coordinates $(\Delta\eta,\Delta\phi)$. These metrics are directly related to the properties of particles, which can guide us to classify the particles behind the jets. To test the impacts of complex features on jet tagging, we also employ the mixed ones $(\log E, \log p_T)$ and $(\log E, \log p_T, \Delta\eta,\Delta\phi)$ to order clusters. From the perspective of graph representation, the nodes are sorted from these components of features respectively, to construct the Haar basis  $\{\boldsymbol{\phi}_l^{(0)}\}_{l=1}^{N_{0}}$ of the original graph $\mathcal{G}_0$ and the corresponding Haar matrix $\tilde{\Phi}^{(0)}\in R^{N_{0}\times N_{0}}$. Then the compressive Haar basis matrix $\Phi_0\in R^{N_{1}\times N_{0}}$ for $N_1<N_0$ leading to the next chain layer $\mathcal{G}_1$ can be directly addressed.
 
\begin{figure}[h]
\includegraphics[scale=0.26]{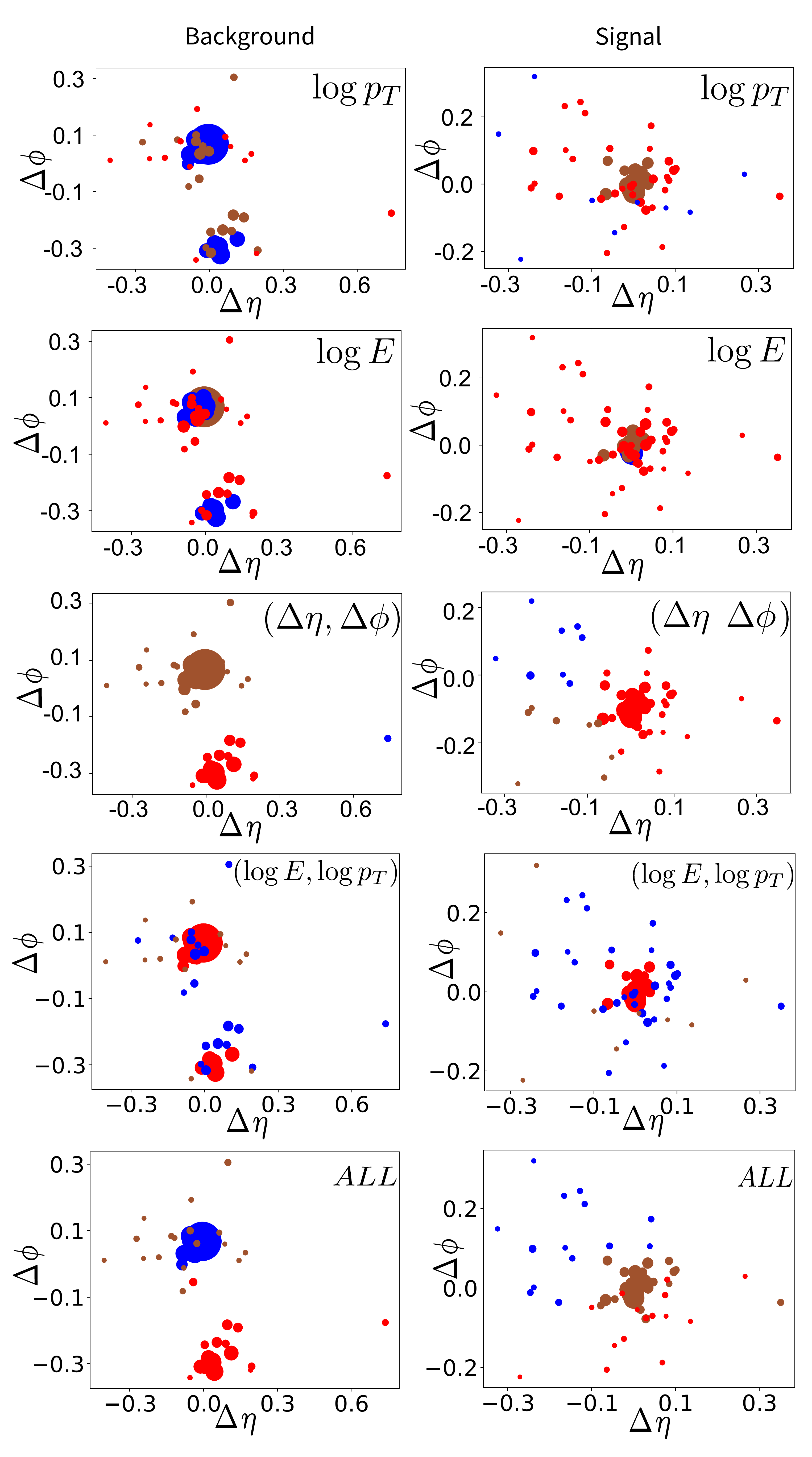}\caption{Example of clustering images for background (gluon) and signal (quark) in columns, with the clustering by $\log E$, $\log p_{T}$, $(\Delta\eta,\Delta\phi)$, $(\log E, \log p_T)$ and ``ALL'' for $(\log E, \log p_T, \Delta\eta,\Delta\phi)$ in rows, respectively. The size of nodes is proportional to $E$, and the nodes owning the same color belong to the same cluster.}
\label{fig:G_P_cluster}
\end{figure}

 In detail, we only keep the information of the top 100 particles in order of momentum for each data of events, for which the excessive particles are not considered or the original data is retained if insufficient~\cite{Moreno:2020up}. The particles in each event is clustered by the $k$-means method~\cite{pakhira2014linear} for $\log E$, $\log p_{T}$ ,$(\Delta\eta,\Delta\phi)$, $(\log E, \log p_T)$ and $(\log E, \log p_T, \Delta\eta,\Delta\phi)$ respectively. The label information of particle features can be obtained by the clustering, and the different structures of $\Phi_0$ can be achieved from the different cluster labels by the three kinds of sorting, as shown in Fig.~\ref{fig:G_P_cluster}. The number of labels in each event is determined by the number of particles and the pooling rate. In the actual process, we first normalize the variable which carries the particle feature to $[0,1]$, and cluster the particles in each event separately. The number of labels is set to particle number in each event times pooling rate. Particles in different events are assigned to all or part of the labels according to the number and information distribution of particles. The pooling rate is set to $0.6$, and also, we will discuss the impact of different pooling rates on the results in Subsection~\ref{Results}. The $k$-means clustering on $500$ events separately costs $17s$.

\begin{figure}
\centering
\includegraphics[scale=0.27]{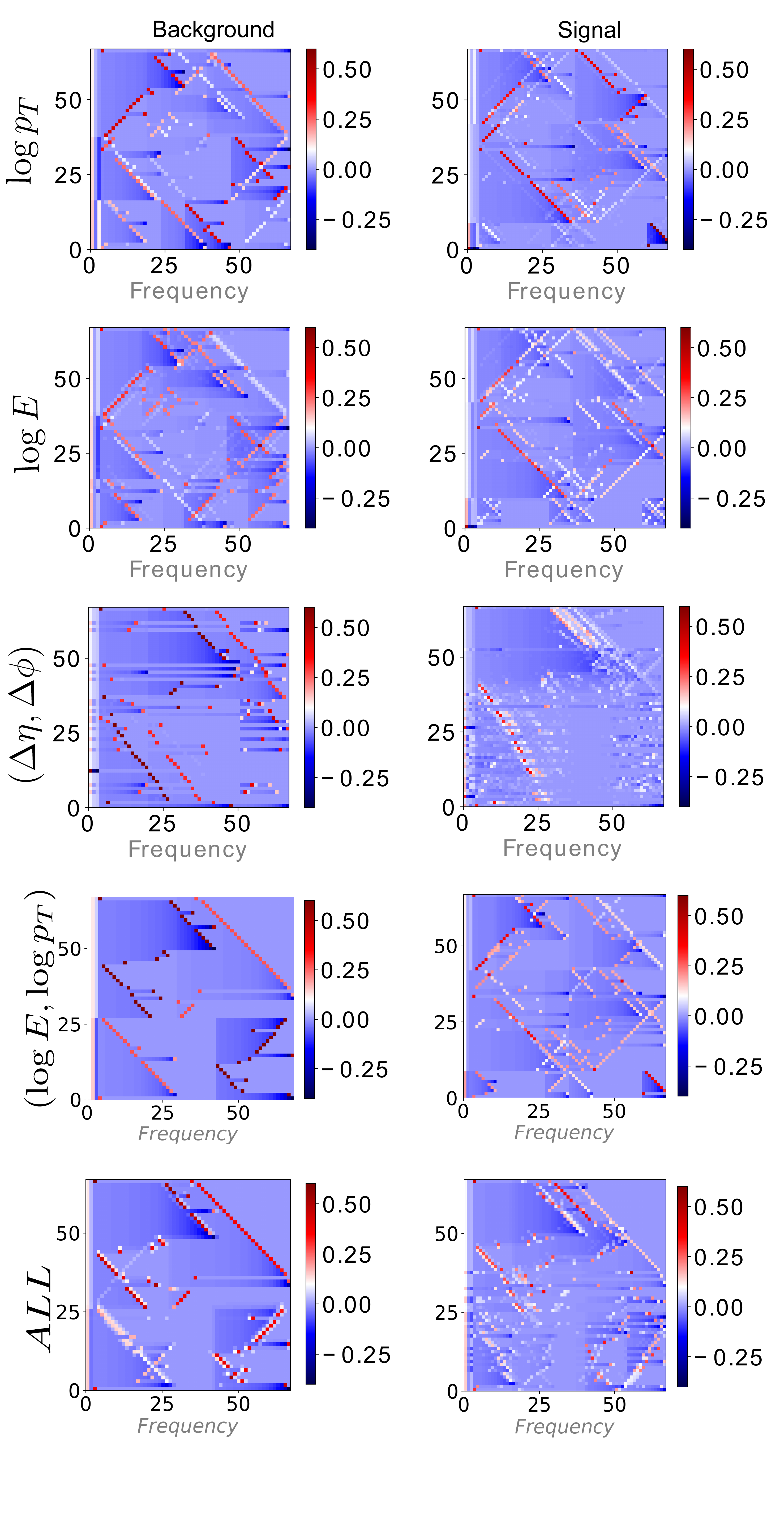}
\caption{Example images of the full frequency Haar bases $\tilde{\Phi}$ for background (gluon) and signal (quark) by $\log E$, $\log p_{T}$, $(\Delta\eta,\Delta\phi)$, $(\log E, \log p_T)$ and ``ALL'' for $(\log E, \log p_T, \Delta\eta,\Delta\phi)$. Here we show  the $67\times 67$ Haar matrices selected from the first $67$ particles in a jet event sorted by $p_T$ with $k = 3$ (clusters) for $100$ times average. The abscissa from 0 to $67$ is the frequency from low to high. The size of the matrix element values is represented by different colors.}
\label{fig:pool_base}
\end{figure}

Now, particle features are transformed into the full frequency Haar base $\tilde{\Phi}_0$ by calculating a fixed $\boldsymbol{\phi}^{(0)}$ with the clustering information, and each $\boldsymbol{\phi}^{(0)}$  represents different frequency in spectral space~\cite{wang2019haarpooling}. The low-frequency coefficients have the local information and the high-frequency coefficients contain the fine details of the observable space. It means in the low frequency space particles with the same clustering label have close value
of elements in Haar matrix, but this does not happen in high-frequency space. From that $\tilde{\Phi}_0$ can learn information of different frequency from the ordering of $\log E$, $\log p_{T}$, $(\Delta\eta,\Delta\phi)$, $(\log E, \log p_T)$ and $(\log E, \log p_T, \Delta\eta,\Delta\phi)$, respectively. In Fig.~\ref{fig:pool_base}, the values of low-frequency coefficients obtained by $\log p_{T}$ and $\log E$ show the information of local clusters as the ``square-like''  structure, and it also happens in the mixed ordering of $(\log E, \log p_T)$. However, this structure disappears in  the low-frequency coefficients obtained by $(\Delta\eta,\Delta\phi)$ and $(\log E, \log p_T, \Delta\eta,\Delta\phi)$ in the space of $\log p_T$, since $\log p_T$ nearly has no relation to $(\Delta\eta,\Delta\phi)$. This indicates that the information of $\tilde{\Phi}_0$ is more complete by choosing less related features for clustering. The compressive Haar basis matrix $\Phi_{j}$ is dynamically updated with the set of $k$-means clustering labels which are all related to the order of input nodes, so the pooling operator is also permutation invariant. 

From the perspective of graph pooling, this labeling of clusters is a process of dividing nodes $v^{(0)}$ in $\mathcal{G}_0$ into different groups with certain patterns, and then map all their information to a ``supernode'', respectively. These ``supernodes'' constitute the nodes $v^{(1)}$ of $\mathcal{G}_1$ for $\mathcal{G}_{0\to 1}$, and with the chain process the number of nodes in each layer is decreasing until $\mathcal{G}_J$ for $N_J = 1$. Since graph pooling is a process of extracting feature and compressing information, too much coarsening would lead to a huge loss of detailed information from $\mathcal{G}_0$.  In our study, pooling was performed only twice as $\mathcal{G}_{0\to 2}$: In the mapping of $\mathcal{G}_{0\to1}$ we used the label information of clustering to construct $\Phi_0$, and for $\mathcal{G}_{1\to 2}$ only the sets of partitions selected by cluster centers are adopted.
 
 \begin{figure*}[htbp]
  \centering
    \includegraphics[width=1\textwidth]{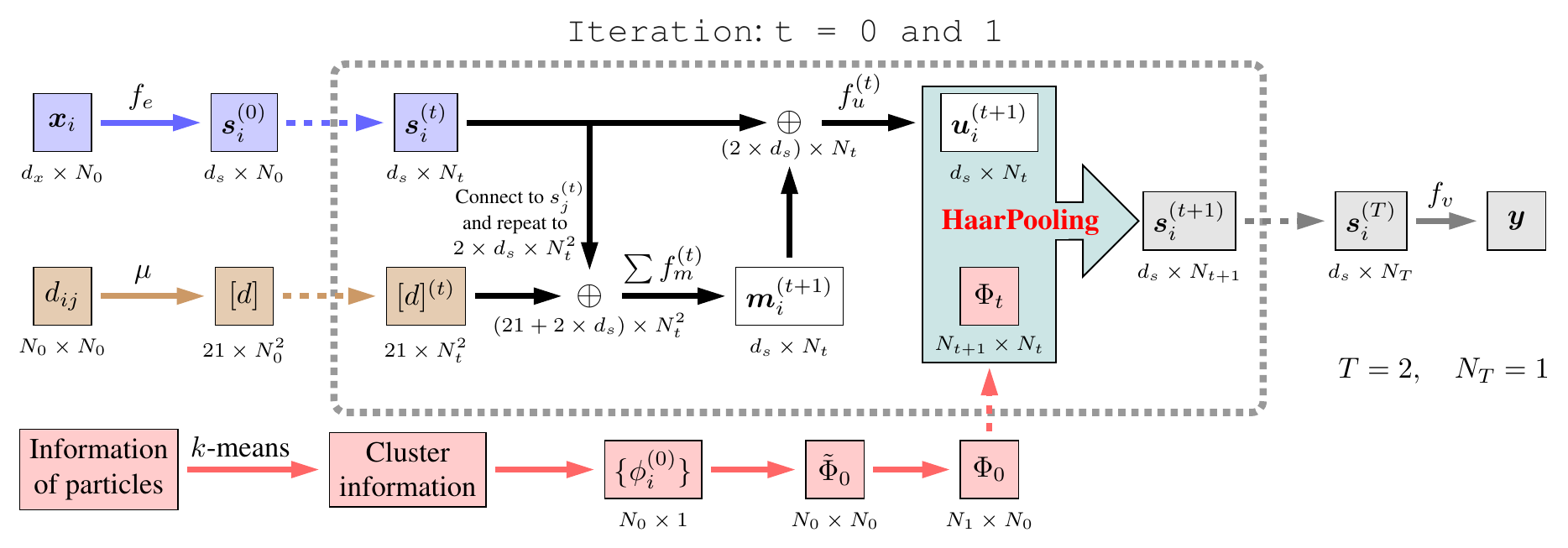}
    \caption{The flow chart illustrating the framework of HMPNet. The blue and brown paths are the input flows of the node weight $\boldsymbol{x}_{i}$ and edge weight $d_{ij}$. The red path represents the process to form the compressive Haar basis $\boldsymbol{\phi}$. The part in large grey dotted box is for the iteration steps from $t =0$ to $1$. The grey path is the output of the iteration area. The labels below each quantity indicates its dimensions.}
    \label{fig:arch}
\end{figure*}
 
\subsection{The HMPNet structure}\label{NetFlow}

The structure of our HMPNet algorithm is shown in Fig.~\ref{fig:arch}, as a process of the MPNN combined with the HaarPooling. The input data is the node weights $\boldsymbol{x}_{i} \in  R^{d_{x}\times N_0}$ from Section \ref{GrapRep}, which is $d_x = 10$ and $N_0 = 30$ in our study. To embed $\boldsymbol{x}_{i}$ in $\boldsymbol{s}_{i}^{(0)} \in R^{d_{s}\times N_0}$ for $d_s = 40$ as Eq.~(\ref{s_i}), the embedding function of NN is given by
\begin{equation}
f_{e}(\boldsymbol{x}_{i}) = relu(W_{e}\boldsymbol{x}_{i} + \boldsymbol{b}_{e}),
\end{equation}
where $W_{e}$ and  $\boldsymbol{b}_{e}$ are hyper-parameters of NN. $relu$ is the Linear rectification function for activation. As a part of the MPNN, the initial information of $\boldsymbol{s}_{i}^{(0)}$ is continuously updated and passed to  $\boldsymbol{s}_{i}^{(t)}$, which is an iterative process from $0$ to $T$. Combined with the HaarPooling from $\mathcal{G}_{0\to J}$, we keep the iteration steps of MPNN consistent with the pooling layers of graph chain, i.e., $J=T=2$, as the steps inside the large grey dotted box of Fig.~\ref{fig:arch}.

For  $\boldsymbol{s}_{i}^{(t)}$, $t=0$ and $1$, the message vector $\boldsymbol{m}_{i}^{(t)}$ of Eq.~(\ref{m_i}) is achieved by the message passing function
\begin{equation}
f_{m}^{(t)}(\boldsymbol{s}_{i},{\boldsymbol{s}_{j},}d) = relu(W_{m}^{(t)}(\boldsymbol{s}_{i}\oplus{\boldsymbol{s}_{j}\oplus}[d]) + \boldsymbol{b}_{m}^{(t)}),
\end{equation}
with hyper-parameters $W_{m}^{(t)}$ and $\boldsymbol{b}_{m}^{(t)}$. Here $\oplus$ means vector concatenation. Here $\boldsymbol{s}_{i}$ is connected to $\boldsymbol{s}_{j}$ so that it can learn information from both sides of an edge in graph representation. As $d_{ij}$ is a real number, to keep the same dimension as $\boldsymbol{s}_{i}$ for passing message easily, it is better to map $d_{ij}$ into a vector $[d]$ by an non-normalized Gaussian basis
\begin{equation}
[d]:(\boldsymbol{d}_{ij})_{k} = e^\frac{(d_{ij} - \mu_{k})^{2}}{2 \sigma^{2}}.
\end{equation}
Here the mean value $\mu_{k}$ is chosen from $[0,5]$ as a uniform distribution: $\mu_{1} = 0,\mu_{2} = 0.25,\cdots,\mu_{20} = 4.75,\mu_{21} = 5$, $\sigma = 0.25$, so $\boldsymbol{(d}_{ij})_{k}$ is a 21-dimensional vector with component information of $N_0\times N_0$. For each iteration $t$, it is updated as $[d]^{(t)}$ of $21\times N_t^2 $. This step is borrowed from the idea of radio basis function (RBF) networks and has been shown to give better results than using $d_{ij}$~\cite{Abdughani:2019uk}. It should be noted that to ensure the uniformity of dimensions, $\boldsymbol{s}_{i}^{(t)}$ is repeatedly referred to as the dimension of $d_s \times N_t^2$. And with the connection 
$\boldsymbol{s}_{i}^{(t)} \oplus \boldsymbol{s}_{j}^{(t)}$, it turns to $2 \times d_s \times N_t^2$ for an extra $d_s$ dimension of $\boldsymbol{s}_{j}^{(t)}$. By the summation of $f_{m}^{(t)}$ in Eq.~(\ref{m_i}), the dimension of $\boldsymbol{m}_{i}^{(t)}$ is back to $d_s\times N_t$.

The update process is the same as in Eq.~(\ref{s_vector}), but the output is not directly used as $\boldsymbol{s}_{i}^{(t+1)}$, which is represented by $\boldsymbol{u}_{i}^{(t+1)}$ now. The update function can be expressed as
\begin{equation}
f_{u}^{(t)}(\boldsymbol{s}_{i},\boldsymbol{m}) = relu(W_{u}^{(t)}(\boldsymbol{s}_{i} \oplus \boldsymbol{m}) + \boldsymbol{b}_{u}^{(t)}),
\label{f_u}
\end{equation}
of hyper-parameters $W_{u}^{(t)}$ and $\boldsymbol{b}_{u}^{(t)}$. Here $\boldsymbol{u}_{i}^{(t+1)}=f_{u}^{(t)} \in R^{N_{t} \times d_s}$ is set as the input feature array for Haar Pooling:
\begin{equation}
\boldsymbol{s}_{i}^{(t+1)} = \Phi_{t}^{T}\boldsymbol{u}_{i}^{(t+1)}, \quad j=0 \text{ and } 1,
\label{Haar_S}
\end{equation} 
where the output $\boldsymbol{s}_{i}^{(t+1)} \in R^{N_{t+1} \times d_s}$ and $N_{t} > N_{t+1}$. To guide the eye, the process to address the compressive Haar matrix $\Phi_{t}$ from particle features (see Section \ref{Cluster}) is also drawn as the red path in Fig.~\ref{fig:arch}. An iterative process $t$ ends here and the updated $\boldsymbol{s}_{i}^{(t+1)}$ is regarded as the new input for the $(t+1)$-th iteration. To optimize this update process, we also use the technique of skip connection~\cite{he2016deep}.

For the last iteration $T=2$,  similar to Eq.~(\ref{y_score}), the signal such as probability $y$ is given by the voting function
\begin{equation}
f_{v}(\boldsymbol{s}_{i}^{(T)}) = sigmoid(W_{v}\boldsymbol{s}_{i}^{(T)} + \boldsymbol{b}_{v}),
\label{f_v}
\end{equation}	
where $W_{v}$ and $\boldsymbol{b}_{v}$ are hyper-paremeters of NN. The $sigmoid$ refers to the activate sigmoid function. Different from Eq.~(\ref{y_score}), the summation is not required here because $\boldsymbol{s}_{i}^{(T)}$ is a vector of $d_s$ components at the end of graph pooling for $\mathcal{G}_T$ with $N_T=1$:
\begin{equation}
y = f_{v}(\boldsymbol{s}_i^{(T)}).
\label{y_fin}	
\end{equation}
Here we also consider event selection efficiency $\varepsilon$ by selecting events with a specific cut threshold $\theta_{y}$ and only events with $y > \theta_y$ are singled out\cite{Abdughani:2019uk}. Our hierarchical approach extracts local information of observable space at different scales.

For each training epoch, we adopted binary-cross-entropy as the loss function for optimization. The ML process is shared for all nodes, and the output does not change with the permutation of input sorted nodes. Our graph feature matrix $\boldsymbol{s}_{i}^{(t)}$  is not fixed but dynamically updated after each layer of the network.

\section{Results and Discussion}

\subsection{Data and settings}\label{Data}

Quark-gluon tagging, aiming to distinguish jets initiated by quarks (signal) and gluons (background), is an important HEP focus related to search for new physics at the LHC. We use the dataset in Ref.~\cite{Komiske:2019um}, to evaluate the performance of the HMPNet. The quark and gluon jets are generated with \textsf{\small PYTHIA 8.226}\cite{Sjostrand:2006vy,Sjostrand:2015vy} with $Z$ decaying to neutrinos, in which $Z(\to v\bar {v}) + (u,d,s)$ as singal jet and $Z(\to v\bar {v}) + g$ is background jet, at $\sqrt{s} = 14$ TeV. The \textsf{\small FastJet 3.3.0} \cite{Cacciari:2012us} is used to cluster final-state non-neutrino particles into $R = 0.4$ anti$-k_{T}$ jets\cite{Cacciari:2008wr}. Only the jets with transverse momentum $p_{T}\in [ 500,550]$ GeV and rapidity $|y_{\text{rap}}| < 2.0$ are considered. No detector simulation is performed here. We follow the recommended splitting dataset to $1.6M/200k/200k$ events, for training, testing and evaluation of the method respectively. The PID information is also used for our jet tagging as the last four components of $\boldsymbol{x_i}$ in Table \ref{tab:input-features}.

The HMPNets is implemented in the open-source DL framework PyTorch 1.8.0 with TensorFlow 2.3.0, so as to compare the MPNN in Ref.~\cite{Abdughani:2019uk}. They were all trained on two NVIDIA 2080 Ti GPUs in parallel.
The Adam optimizer~\cite{Adam} is used to speed up the training process of the GNNs.
The batch size is set to $160$ and $GradualWarmupScheduler$~\cite{2017Accurate} is chosen to optimize the training process. In detail, a warm-up period lasting $4$ epochs is applied before reaching the initial learning rate $1\times 10^{-3}$, and a $CosineAnnealingWarmRestarts$ learning rate schedule by a factor of $2$ at every restart~\cite{2016SGDR} is adopted for the next $28$ epochs. Finally, an learning rate of exponential decay, of exponent 0.5, is used for the last $3$ epochs, similarly as in~\cite{Gong:2022wk}. So the total number of training epochs is 35.

\subsection{Results}\label{Results}

\begin{figure}
\includegraphics[scale=0.35]{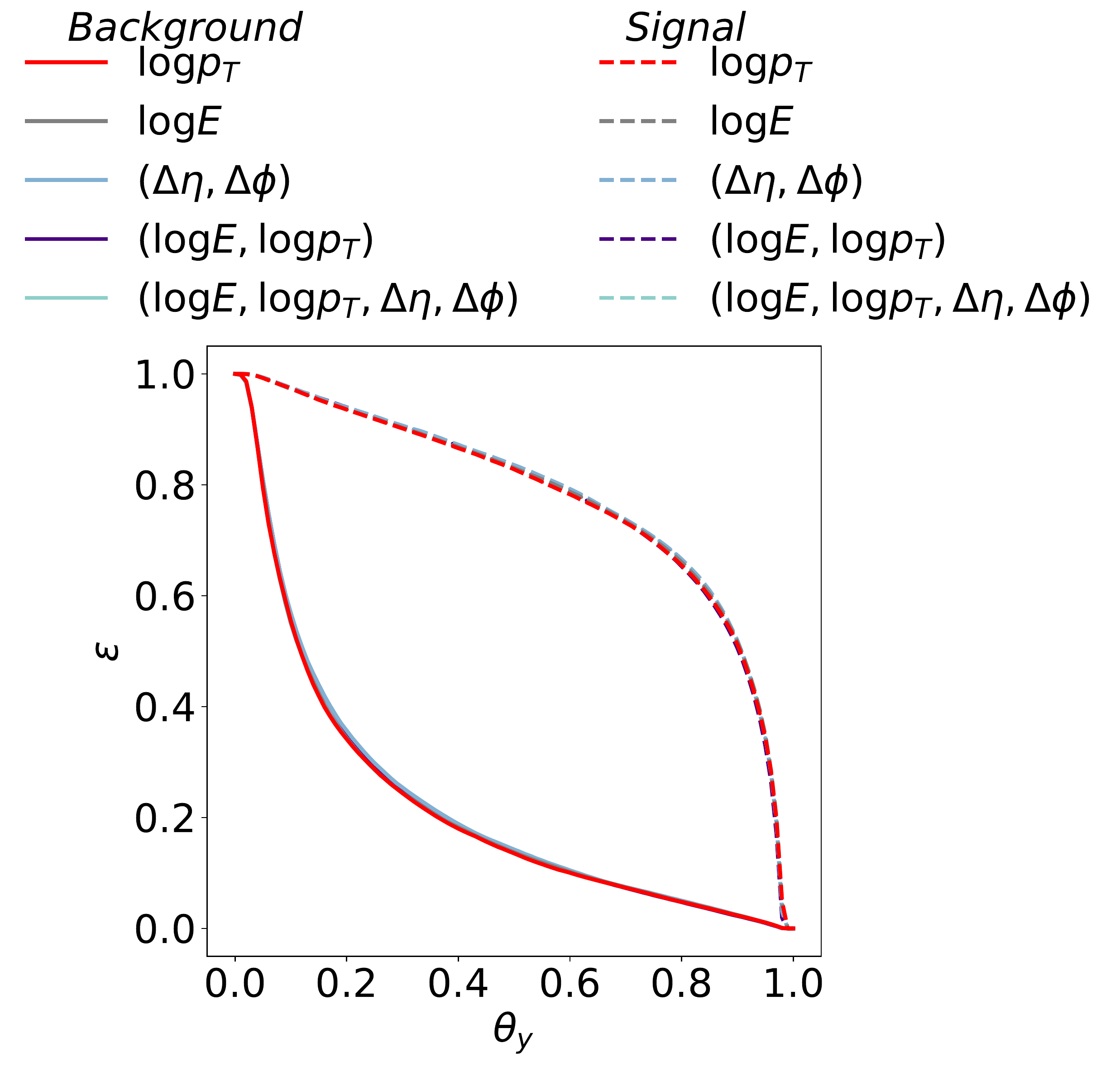}
\caption{The selection efficiency curves of HMPNet with particle features $\log p_{T}$, $\log E$, $(\Delta\eta,\Delta\phi)$, $(\log E, \log p_T)$ and $(\log E, \log p_T, \Delta\eta,\Delta\phi)$. Selection efficiency and cut threshold are denoted as $\varepsilon$ and $\theta_{y}$, respectively.}
\label{fig:eff_diff}
\end{figure}

For simplicity, we denote the results of HMPNets with the Haar base information from the ordering of $\log p_{T}$, $\log E$, $(\Delta\eta,\Delta\phi)$, $(\log E, \log p_T)$ and $(\log E, \log p_T, \Delta\eta,\Delta\phi)$, respectively. In Fig.~\ref{fig:eff_diff}, the selection efficiency curves of the results show that all of them can distinguish the background and the signal well. However, it is difficult to tell the difference of the results by the naked eyes, so we compare them across metrics lists in Table.~\ref{tab.diff_mpnn}. The receiver operating characteristic (ROC) is obtained from the true positive rates $\varepsilon_{S}$ and false positive rates $\varepsilon_{B}$ with a changing decision threshold. Usually for two curves of ROC, their difference can be evaluated by the area under the ROC curve (AUC). Another important metric is the background rejection at a certain signal efficiency $R_{\varepsilon_{s}} = 1/ \varepsilon_{B}$ @ $\varepsilon_{s}$ for $R_{\varepsilon_{s}=50\%}$ and $R_{\varepsilon_{s}=30\%}$.  From Table.~\ref{tab.diff_mpnn}, it is obvious that the performance of $\log P_T$ is the best among all the metrics, which indicates that $P_T$ is probably the most relevant particle feature for jet tagging. From the column of accuracy, it seems that most particle features do not show much different influence on the results, except that the AUG and $R_{\varepsilon_{s}=50\%}$ of $(\Delta\eta, \Delta\phi)$ are not as good as other orderings. We hence conjecture that the coordinate information of $(\Delta\eta, \Delta\phi)$ may be less relevant to particle classification, whose joining only makes the results more skewed. From the results of $(\log E, \log p_T)$ and $(\log E, \log p_T, \Delta\eta,\Delta\phi)$, it shows that adding more information is not helping, but may interfere with the learning of NN. Therefore, choosing appropriate information for the Haar basis is very crucial. It is worth noting that we also give the results of MPNN without HaarPooling process in the last low of Table~\ref{tab.diff_mpnn}. For each metric the MPNN result is worse than any others with additional particle features, which indicates  the HaarPooling process has significantly enhanced the power of NN to extract and learn particle features.

\begin{table}
\caption{Performance comparison on the quark-gluon tagging of HMPNet with different particle features. The last row of ``None" is for the results of MPNN without HaarPooling. The uncertainty quoted corresponds to the standard deviation of $R_{\varepsilon_S}$ with a certain $\varepsilon_S$.}
\scalebox{0.9}{
\begin{tabular}{c c c c c}
\hline 
\hline
 Particle feature & Accuracy & AUC & $R_{\varepsilon_{S} = 50\%}$ & $R_{\varepsilon_{S} = 30\%}$ \\
 \hline
 $\log p_T$ &$\boldsymbol{0.846}$ & $\boldsymbol{0.9185}$ & $\boldsymbol{45.2\pm 0.3}$ & $\boldsymbol{118.1\pm 1.2}$ \\
 $\log E$ &$0.845$ & $0.9173$ & $43.2\pm 0.2$ & $115.4\pm 1.4$ \\
 $(\Delta \eta, \Delta \phi)$ & $0.844$ & $0.9166$ & $42.1\pm 0.2$ & $113.6\pm 1.3$ \\
 $(\log E, \log p_T)$ & $0.844$  & $0.9169$ & $43.3\pm 0.2$  & $112.1\pm 1.5$\\
 $(\log E, \log p_T, \Delta \eta, \Delta \phi)$ & $0.845$  & $0.9172$ & $44.3\pm 0.4$  & $116.2\pm 1.2$\\
  \hline
 None & $0.839$ & $0.9118$ & $39.3\pm 0.2$ & $98.0\pm 1.4$ \\
 \hline
 \hline
\end{tabular}}
\label{tab.diff_mpnn}
\end{table}

\begin{figure}
\includegraphics[scale=0.45]{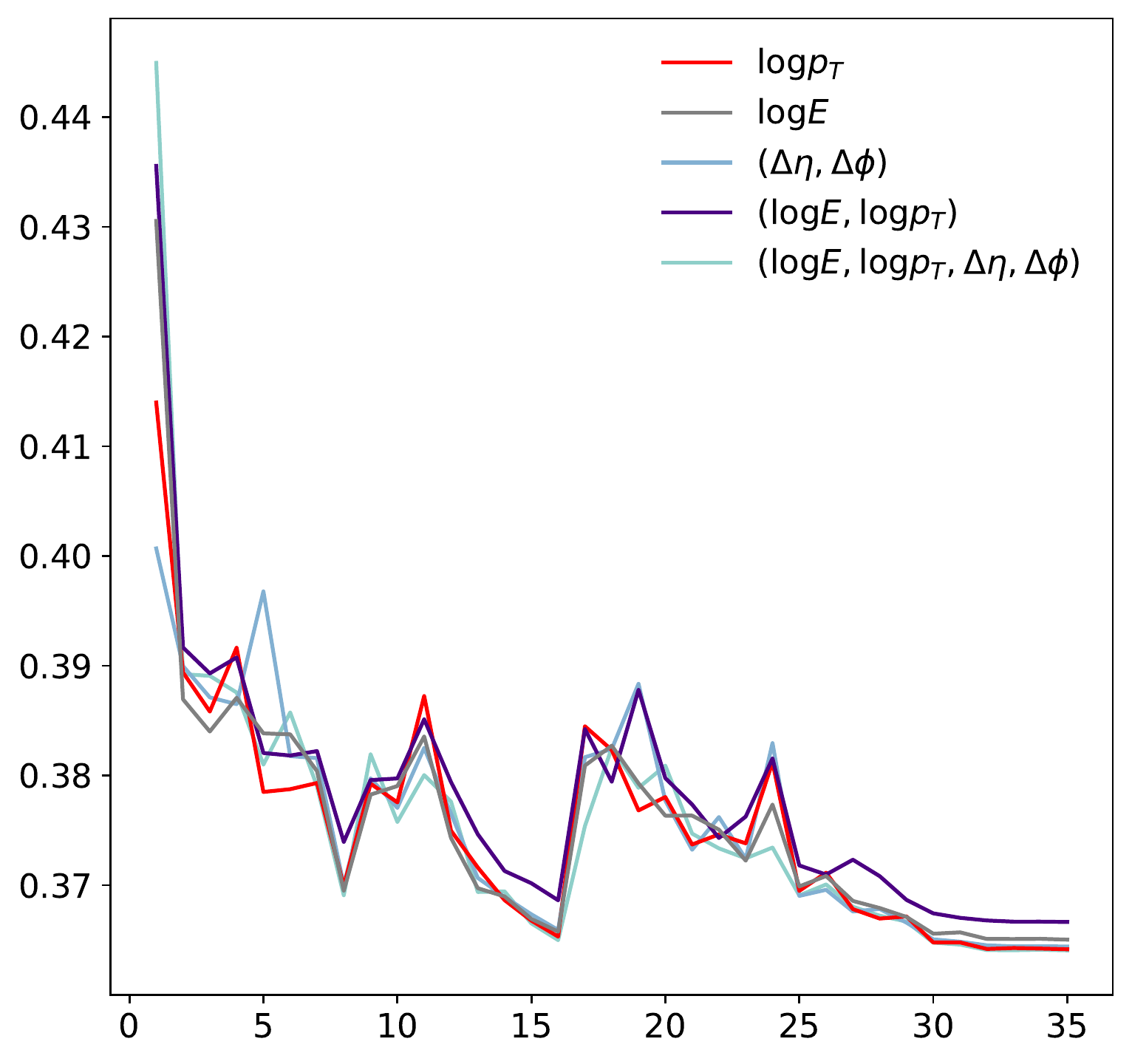}
\caption{The loss history of HMPNet with particle features $\log p_{T}$, $\log E$, $(\Delta\eta,\Delta\phi)$, $(\log E, \log p_T)$ and $(\log E, \log p_T, \Delta\eta,\Delta\phi)$.}
\label{fig:loss}
\end{figure}

The loss history of HMPNet with different particle features is shown in Fig.~\ref{fig:loss}. All the loss functions converge smoothly at epoch $=35$, so that the results are relatively stable and convincing. We take the average of five independent runs, and by tests the results are very robust with nearly negligible deviation when $\varepsilon_S>0.2$. We also give the accuracy and AUC of HMPNet with particle features $\log p_{T}$, $\log E$, $(\Delta\eta,\Delta\phi)$, $(\log E, \log p_T)$ and $(\log E, \log p_T, \Delta\eta,\Delta\phi)$ by different pooling rates of $0.4, 0.6$ and $0.8$ in Table~\ref{tab.diff_rate}. From Table~\ref{tab.diff_rate} one can see the results remain nearly unchanged with the pooling rate when it is larger than $0.4$, which also indicates the results of HMPNet are stable and only affected by different particle features.

\begin{table}[h]
\centering
\caption{Comparison of the quark-gluon classification performance of HMPNet results with different pooling ratios, via ACC and AUC. The uncertainty quoted corresponds to the standard deviation of five trainings with different random weight initialisations. If the uncertainty is not quoted then the variation is negligible compared to the expected value.}
\scalebox{0.9}{
\begin{tabular}{|c| c c c|}
\hline
Particle feature & Pooling rate & Accuracy & AUC \\
\hline
\multirow{3}{*}{$\log p_T$} & $0.4$ & $0.845\pm 0.002$ & $0.9180\pm 0.0012$\\
& $0.6$ & $0.846\pm 0.001$ & $0.9185\pm 0.0007$\\
& $0.8$ & $0.846\pm 0.002$ & $0.9179\pm 0.0008$\\
\hline
\multirow{3}{*}{$\log E$} & $0.4$ & $0.846\pm 0.001$ & $0.9170\pm 0.0013$\\
& $0.6$  &$0.845\pm 0.001$ & $0.9173\pm 0.0011$\\
& $0.8$ & $0.845\pm 0.001$ & $0.9178\pm 0.0009$\\
\hline
\multirow{3}{*}{$(\Delta \eta, \Delta \phi)$} & $0.4$ & $0.844\pm 0.001$ & $0.9158\pm 0.0011$\\
& $0.6$ & $0.844\pm 0.001$ & $0.9166\pm 0.0012$\\
& $0.8$ & $0.845\pm 0.001$ & $0.9162\pm 0.0019$\\
\hline
\multirow{3}{*}{$(\log E, \log p_T)$} & $0.4$ & $0.844\pm 0.001$ & $0.9163\pm 0.0013$\\
& $0.6$ & $0.844\pm 0.002$  & $0.9169\pm 0.0008$\\
& $0.8$ & $0.845\pm 0.001$ & $0.9167\pm 0.0011$\\
\hline
\multirow{3}{*}{$(\log E, \log p_T, \Delta \eta, \Delta \phi)$} & $0.4$ & $0.845\pm 0.002$ & $0.9171\pm 0.0018$\\
& $0.6$ & $0.845\pm 0.001$  & $0.9172\pm 0.0010$\\
& $0.8$ & $0.846\pm 0.001$ & $0.9177\pm 0.0008$\\
\hline
\end{tabular}}
\label{tab.diff_rate}
\end{table}

\begin{figure}
\includegraphics[scale=0.35]{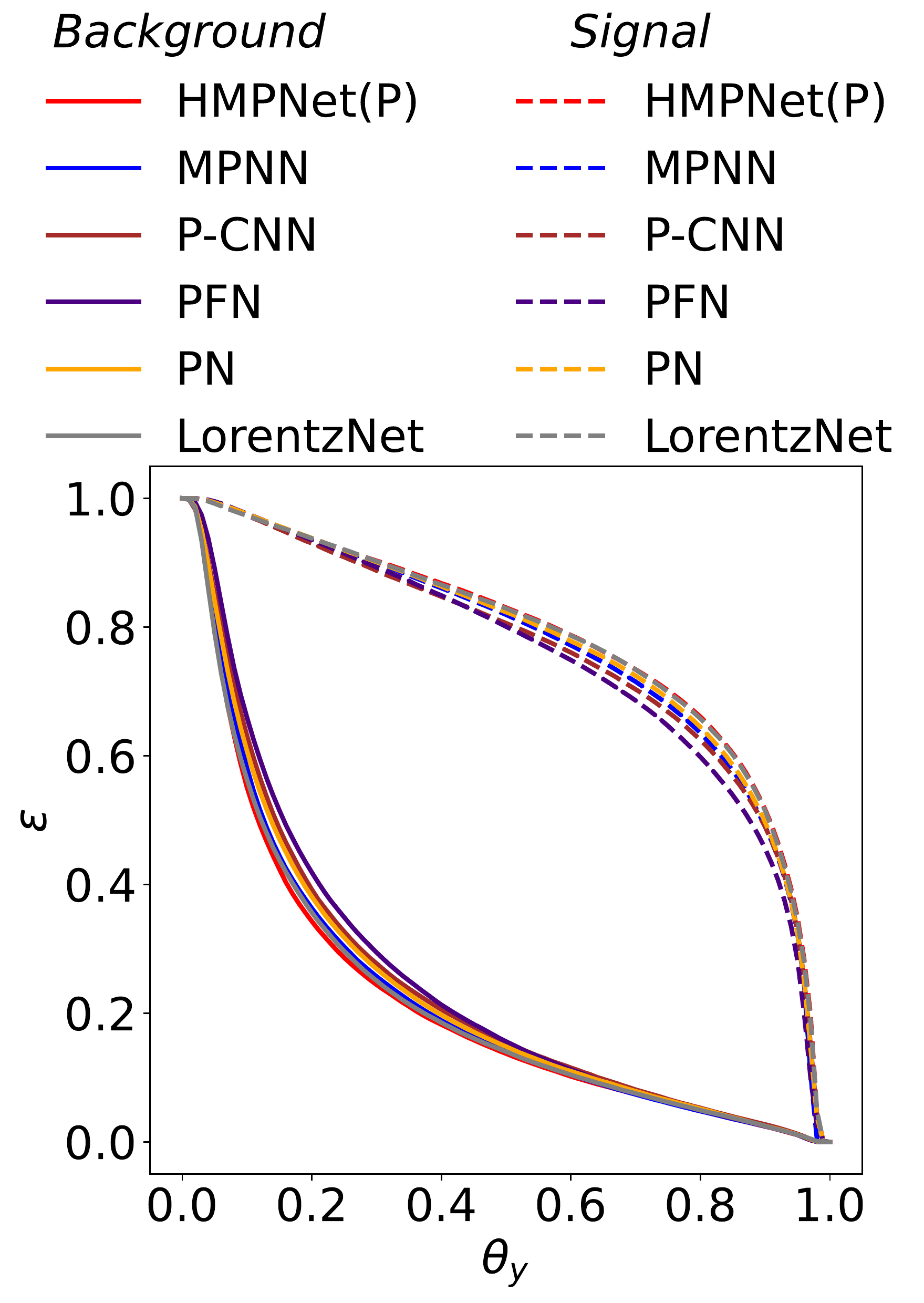}%{QG_PRED.pdf}
\caption{The selection efficiency curves of models. Selection efficiency and cut threshold are denoted as $\varepsilon$ and $\theta_{y}$, respectively.}\label{fig:qg-pred}
\end{figure}
 
\begin{figure}[htbp]
\centering
\subfigure[ROC]{
\label{ROC}
\includegraphics[width=0.2235\textwidth]{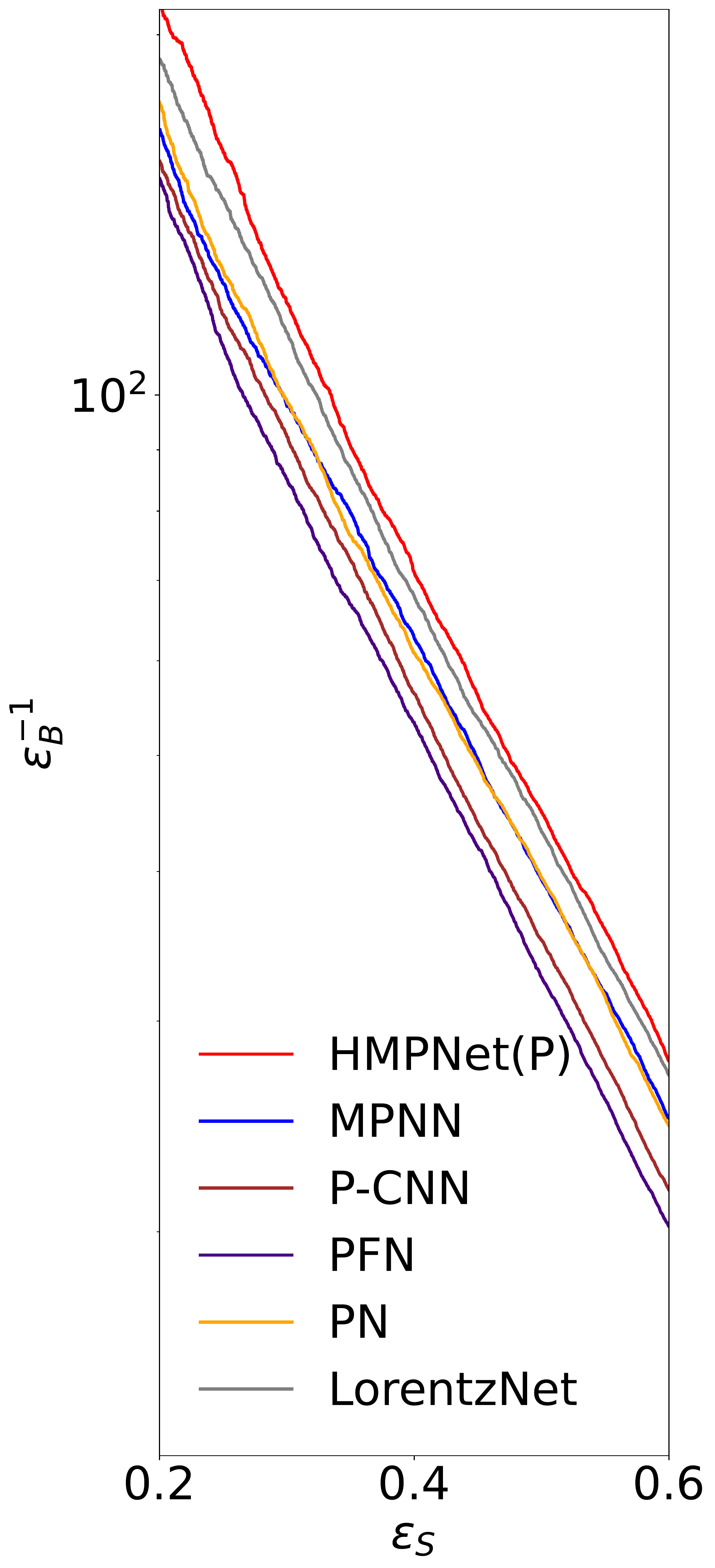}}
\subfigure[SI]{
\label{SI}
\includegraphics[width=0.22\textwidth]{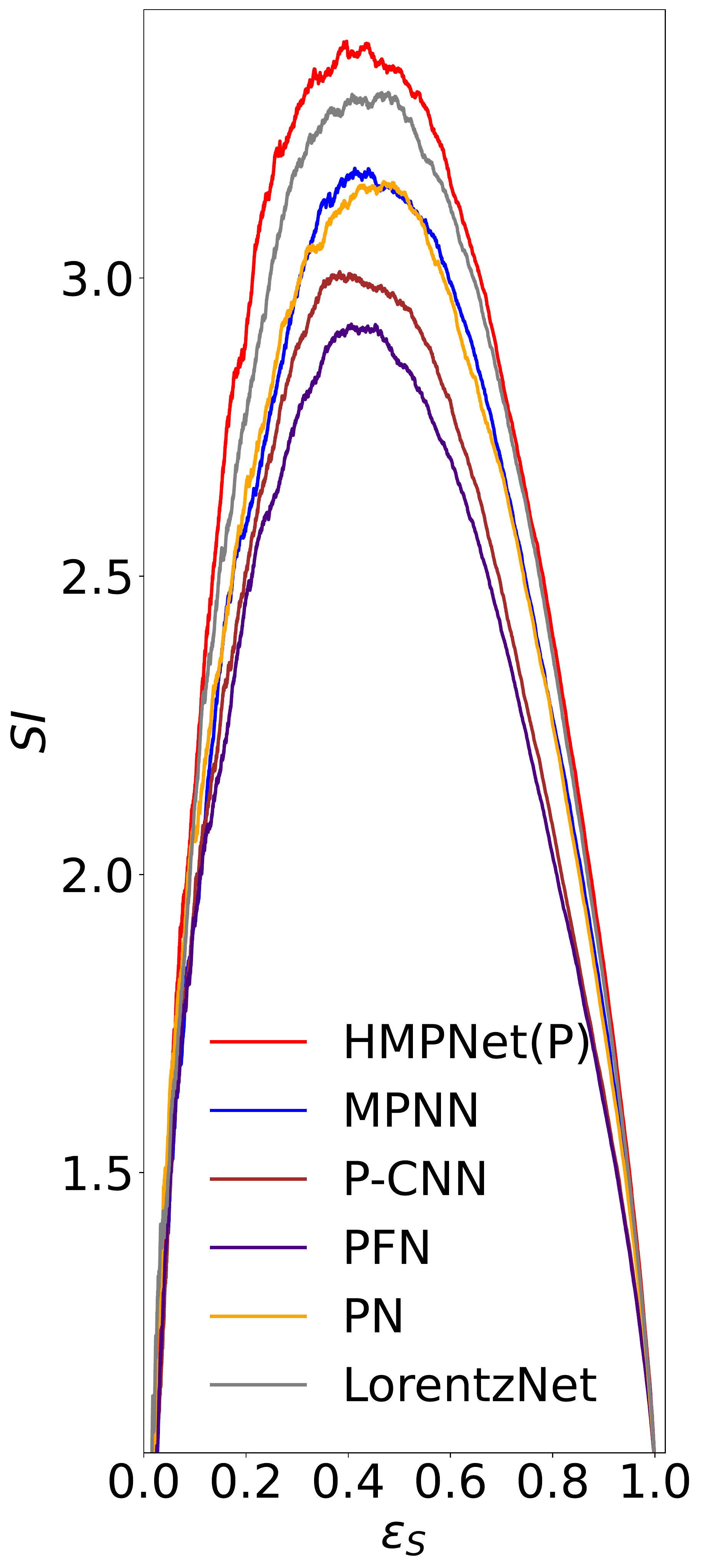}}
\caption{(a) The ROC curves of models with $\varepsilon_{S}$ and $\varepsilon_{B}$. (b) The SI curves of models with $SI = \varepsilon_{S}/ \sqrt{\varepsilon_{B}}$ and $\varepsilon_{S}$.}
\label{fig:ROC_and_SI}
\end{figure}

Furthermore, we compare the HMPNet results ordered by particle feature $\log p_T$ as HMPNet(P) to previous studies: P-CNN~\cite{cms2017boosted},PFN~\cite{Komiske:2019um},PN~\cite{qu2020jet} and LorentzNet~\cite{Gong:2022wk}. Fig.~\ref{fig:qg-pred} shows that the selection efficiency curves of all the methods are close to one others. To see more significant differences, we give the curves of ROC in Fig.~\ref{ROC} and the Significance Improvement (SI) curves evaluated by $SI = \varepsilon_{S}/ \sqrt{\varepsilon_{B}}$~\cite{Gallicchio:2013ww} in Fig.~\ref{SI}.
It is obvious that the performance of HMPNet(P) is better than the rest ones within $\varepsilon_s\in[0.2,0.6]$. In detail, the accuracy, AUC and background rejection results are summarized in Table \ref{tab.result_all}. In the columns of the accuracy, AUC and $R_{\varepsilon_{s}=50\%}$, the values of HMPNet(P) are slightly larger ($0.5\%-7\%$) than others. For $R_{\varepsilon_{s}=30\%}$, it is $118.1\pm 1.2$, second only to the $118.4\pm 1.5$ of ABCNet.
These findings fully demonstrate that HMPNet(P) is an outstanding choice of NN algorithm for jet tagging. %\tcr{Add something about FLOPs?}

The evaluation time per batch, the number of trainable parameters and FLOPs of PN, LorentzNet, MPNN and HMPNet on the same GPU cluster with batch size $160$  are given in Table~\ref{tab.info}. The HMPNet and MPNN require muss less number of trainable parameters than PN and LorentzNet, at the same level or even better on FLOPs. Since a fine-designed
pooling operator in GNN can reduce the size of graphs~\cite{Grattarola2022}, the HMPNet does not reduce computational efficiency, compared to the MPNN. The evaluation time on GPUs of HMPNet and MPNN are close, but $30\%$ smaller than LorentzNet.

\begin{table}
\caption{Performance comparison on the quark-gluon tagging of the algorithms. The uncertainty quoted corresponds to the standard deviation of $R_{\varepsilon_S}$ with a certain $\varepsilon_S$. The largest values of each column is highlighted in bold. }
\begin{tabular}{c c c c c}
\hline 
\hline
 Model & Accuracy & AUC & $R_{\varepsilon_{S} = 50\%}$ & $R_{\varepsilon_{S} = 30\%}$ \\
 \hline
 P-CNN~\cite{cms2017boosted} & $0.827$ & $0.9002$ & $34.7$ & $91.0$ \\
 PFN~\cite{Komiske:2019um} & $-$ & $0.9005$ & $34.7\pm 0.4$ & $-$ \\
 PN~\cite{qu2020jet} & $0.840$ & $0.9116$ & $39.8\pm 0.2$ & $98.6\pm 1.3$ \\
 ABCNet~\cite{mikuni2020abcnet} & $0.840$  & $0.9126$ & $42.6\pm 0.4$  & $\boldsymbol{118.4\pm 1.5}$\\
 LorentzNet~\cite{Gong:2022wk} & $0.844$  & $0.9156$ & $42.4\pm 0.4$ & $110.2\pm 1.3$ \\
 MPNN & $0.839$ & $0.9118$ & $39.3\pm 0.2$ & $98.0\pm 1.4$ \\
 HMPNet(P) & $\boldsymbol{0.846}$  & $\boldsymbol{0.9185}$ & $\boldsymbol{45.2\pm 0.3}$ & $118.1 \pm 1.2$ \\
 \hline
 \hline
\end{tabular}
\label{tab.result_all}
\end{table}

\begin{table}
\caption{The comparison of evaluation time, the number of trainable parameters and FLOPs. The models are executed on a cluster with NVIDIA 2080 Ti GPUs in parallel.}
\begin{tabular}{c c c c c}
\hline 
\hline
 Model & Evaluation time (ms/batch) & Params & FLOPs\\
 \hline
 PN & $8.38$ & $366k$ & $540M$\\
 LorentzNet & $15.36$  & $224k$ & $658M$\\
 MPNN & $10.08$ & $58k$ & $521 M$\\
 HMPNet & $10.72$  & $58k$ & $419 M$\\
 \hline
 \hline
\end{tabular}
\label{tab.info}
\end{table}

\section{Conclusion}\label{Conclusion}

In this paper, we employ the HMPNet, a method of GNNs to handle quark-gluon tagging. This method combines MPNN with HaarPooling, which embeds additional information from the input of particle features through the compressed Haar basis matrix $\Phi^{(j)}$, in the process of message passing. This additional information increases the richness of the features and the accuracy of information for updating. 
The Haar basis $\{ \boldsymbol{\phi}_l^{(j)}\}_{l=1}^{N_{j}}$ is obtained by clustering the input particle data via $k$-means of features $\log E$, $\log p_{T}$, $(\Delta\eta,\Delta\phi)$, $(\log E, \log p_T)$ and $(\log E, \log p_T, \Delta\eta,\Delta\phi)$, respectively. By analyzing the $\Phi_j$ composed of these five sorting, it can be clearly seen that the information they convey is different in frequency. On one hand, with the features of $\log p_T$ and $\log E$, the Haarpooling shows a significant improvement of performance. On the other hand, adding relative coordinates information $(\Delta\eta,\Delta\phi)$ is not very beneficial. We also added mixed features as $(\log E, \log p_T)$ and $(\log E, \log p_T, \Delta\eta,\Delta\phi)$ for test, and their results are not as good as adding $\log p_T$ alone but better than using $(\Delta\eta,\Delta\phi)$, which shows more irrelevant additional information would affect the results on the contrary. This indicates that adding information having strong correlation to particle properties enhance the accuracy of quark-gluon tagging, otherwise the results display more deviation. Of course, compared to the normal MPNN, adding any effective information via HaarPooling can enhance the power to extract features. The results of HMPNet are quite stable and barely change for pooling rate larger than $0.4$. By comparing the results of HMPNet(P) with the quark-gluon tagging results through other algorithms, we show that adding extra information of $\log p_T$ to the HMPNet is very competitive with its great performance. Compared to PN, LorentzNet, and MPNN, the computational efficiency of HMPNet is also impressive.

HaarPooling is not only an operation that compresses the dimension of the graph to extract features, but can play a role in adding extra information in the process of GNNs. From Ref.~\cite{wang2019haarpooling},  we know that HaarPooling can be applied in conjunction with any graph convolution in GNNs, so the ML processes of similar methods to study HEP problems could be improved by embedding a Haar matrix operation. The choice of clustering variables to define labels of the HaarPooling operation plays an important role in the performance of the algorithm and is wise to test different choices of features when implementing the classifier. This requires in-depth analysis and mastery of the internal relationship between input data and expected results.

.

\section*{Acknowledgement}

We thank P\'app Gabor, Jie Ren and Jin Min Yang for their helpful suggestions, and Shengfeng Deng for helping check the manuscript. This work was supported in part by the Fun-damental  Research  Funds  for  the  Central  Universities, China (Grant No.  CCNU19QN029), the National Natural Science Foundation of China (Grant No.  11505071,61702207 and 61873104), and the 111 Project 2.0, with Grant No.  BP0820038.

%{\it Note added.} After finishing all the simulations, we learned that a relevant work on the determination of impact parameter in high-energy heavy-ion collisions has been published recently \cite{Mallick:2021wop}. This work, however, used the boosted decision trees, instead of DNN or CNN to recognize the impact parameters, with charged-particle multiplicity and mean transverse momentum as input data.

%\begin{appendix}

%\end{appendix}
%\clearpage
\newpage

\bibliography{refs}

\begin{thebibliography}{73}%
\makeatletter
\providecommand \@ifxundefined [1]{%
 \@ifx{#1\undefined}
}%
\providecommand \@ifnum [1]{%
 \ifnum #1\expandafter \@firstoftwo
 \else \expandafter \@secondoftwo
 \fi
}%
\providecommand \@ifx [1]{%
 \ifx #1\expandafter \@firstoftwo
 \else \expandafter \@secondoftwo
 \fi
}%
\providecommand \natexlab [1]{#1}%
\providecommand \enquote  [1]{``#1''}%
\providecommand \bibnamefont  [1]{#1}%
\providecommand \bibfnamefont [1]{#1}%
\providecommand \citenamefont [1]{#1}%
\providecommand \href@noop [0]{\@secondoftwo}%
\providecommand \href [0]{\begingroup \@sanitize@url \@href}%
\providecommand \@href[1]{\@@startlink{#1}\@@href}%
\providecommand \@@href[1]{\endgroup#1\@@endlink}%
\providecommand \@sanitize@url [0]{\catcode `\\12\catcode `\$12\catcode
  `\&12\catcode `\#12\catcode `\^12\catcode `\_12\catcode `\%12\relax}%
\providecommand \@@startlink[1]{}%
\providecommand \@@endlink[0]{}%
\providecommand \url  [0]{\begingroup\@sanitize@url \@url }%
\providecommand \@url [1]{\endgroup\@href {#1}{\urlprefix }}%
\providecommand \urlprefix  [0]{URL }%
\providecommand \Eprint [0]{\href }%
\providecommand \doibase [0]{http://dx.doi.org/}%
\providecommand \selectlanguage [0]{\@gobble}%
\providecommand \bibinfo  [0]{\@secondoftwo}%
\providecommand \bibfield  [0]{\@secondoftwo}%
\providecommand \translation [1]{[#1]}%
\providecommand \BibitemOpen [0]{}%
\providecommand \bibitemStop [0]{}%
\providecommand \bibitemNoStop [0]{.\EOS\space}%
\providecommand \EOS [0]{\spacefactor3000\relax}%
\providecommand \BibitemShut  [1]{\csname bibitem#1\endcsname}%
\let\auto@bib@innerbib\@empty
%</preamble>
\bibitem [{\citenamefont {Beaudette}(2014)}]{beaudette2014cms}%
  \BibitemOpen
  \bibfield  {author} {\bibinfo {author} {\bibfnamefont {F.}~\bibnamefont
  {Beaudette}},\ }\href@noop {} {\bibfield  {journal} {\bibinfo  {journal}
  {arXiv preprint arXiv:1401.8155}\ } (\bibinfo {year} {2014})}\BibitemShut
  {NoStop}%
\bibitem [{\citenamefont {Sirunyan}\ and\ \citenamefont {et~al. {CMS
  Collaboration}}(2017)}]{Sirunyan_2017}%
  \BibitemOpen
  \bibfield  {author} {\bibinfo {author} {\bibfnamefont {A.}~\bibnamefont
  {Sirunyan}}\ and\ \bibinfo {author} {\bibnamefont {et~al. {CMS
  Collaboration}}},\ }\href {\doibase 10.1088/1748-0221/12/10/p10003}
  {\bibfield  {journal} {\bibinfo  {journal} {Journal of Instrumentation}\
  }\textbf {\bibinfo {volume} {12}},\ \bibinfo {pages} {P10003} (\bibinfo
  {year} {2017})}\BibitemShut {NoStop}%
\bibitem [{\citenamefont {Aaboud}\ and\ \citenamefont {et~al. {ATLAS
  Collaboration}}(2017)}]{Aaboud:2017tm}%
  \BibitemOpen
  \bibfield  {author} {\bibinfo {author} {\bibfnamefont {M.}~\bibnamefont
  {Aaboud}}\ and\ \bibinfo {author} {\bibnamefont {et~al. {ATLAS
  Collaboration}}},\ }\href {\doibase 10.1140/epjc/s10052-017-5031-2}
  {\bibfield  {journal} {\bibinfo  {journal} {The European Physical Journal C}\
  }\textbf {\bibinfo {volume} {77}},\ \bibinfo {pages} {466} (\bibinfo {year}
  {2017})}\BibitemShut {NoStop}%
\bibitem [{\citenamefont {Gallicchio}\ and\ \citenamefont
  {Schwartz}(2011)}]{PhysRevLett.107.172001}%
  \BibitemOpen
  \bibfield  {author} {\bibinfo {author} {\bibfnamefont {J.}~\bibnamefont
  {Gallicchio}}\ and\ \bibinfo {author} {\bibfnamefont {M.~D.}\ \bibnamefont
  {Schwartz}},\ }\href {\doibase 10.1103/PhysRevLett.107.172001} {\bibfield
  {journal} {\bibinfo  {journal} {Phys. Rev. Lett.}\ }\textbf {\bibinfo
  {volume} {107}},\ \bibinfo {pages} {172001} (\bibinfo {year}
  {2011})}\BibitemShut {NoStop}%
\bibitem [{\citenamefont {Larkoski}\ \emph {et~al.}(2014)\citenamefont
  {Larkoski}, \citenamefont {Thaler},\ and\ \citenamefont
  {Waalewijn}}]{Larkoski:2014vc}%
  \BibitemOpen
  \bibfield  {author} {\bibinfo {author} {\bibfnamefont {A.~J.}\ \bibnamefont
  {Larkoski}}, \bibinfo {author} {\bibfnamefont {J.}~\bibnamefont {Thaler}}, \
  and\ \bibinfo {author} {\bibfnamefont {W.~J.}\ \bibnamefont {Waalewijn}},\
  }\href {\doibase 10.1007/JHEP11(2014)129} {\bibfield  {journal} {\bibinfo
  {journal} {Journal of High Energy Physics}\ }\textbf {\bibinfo {volume}
  {2014}},\ \bibinfo {pages} {129} (\bibinfo {year} {2014})}\BibitemShut
  {NoStop}%
\bibitem [{\citenamefont {Bhattacherjee}\ \emph {et~al.}(2015)\citenamefont
  {Bhattacherjee}, \citenamefont {Mukhopadhyay}, \citenamefont {Nojiri},
  \citenamefont {Sakaki},\ and\ \citenamefont {Webber}}]{Bhattacherjee:2015ta}%
  \BibitemOpen
  \bibfield  {author} {\bibinfo {author} {\bibfnamefont {B.}~\bibnamefont
  {Bhattacherjee}}, \bibinfo {author} {\bibfnamefont {S.}~\bibnamefont
  {Mukhopadhyay}}, \bibinfo {author} {\bibfnamefont {M.~M.}\ \bibnamefont
  {Nojiri}}, \bibinfo {author} {\bibfnamefont {Y.}~\bibnamefont {Sakaki}}, \
  and\ \bibinfo {author} {\bibfnamefont {B.~R.}\ \bibnamefont {Webber}},\
  }\href {\doibase 10.1007/JHEP04(2015)131} {\bibfield  {journal} {\bibinfo
  {journal} {Journal of High Energy Physics}\ }\textbf {\bibinfo {volume}
  {2015}},\ \bibinfo {pages} {131} (\bibinfo {year} {2015})}\BibitemShut
  {NoStop}%
\bibitem [{\citenamefont {Ferreira~de Lima}\ \emph {et~al.}(2017)\citenamefont
  {Ferreira~de Lima}, \citenamefont {Petrov}, \citenamefont {Soper},\ and\
  \citenamefont {Spannowsky}}]{PhysRevD.95.034001}%
  \BibitemOpen
  \bibfield  {author} {\bibinfo {author} {\bibfnamefont {D.}~\bibnamefont
  {Ferreira~de Lima}}, \bibinfo {author} {\bibfnamefont {P.}~\bibnamefont
  {Petrov}}, \bibinfo {author} {\bibfnamefont {D.}~\bibnamefont {Soper}}, \
  and\ \bibinfo {author} {\bibfnamefont {M.}~\bibnamefont {Spannowsky}},\
  }\href {\doibase 10.1103/PhysRevD.95.034001} {\bibfield  {journal} {\bibinfo
  {journal} {Phys. Rev. D}\ }\textbf {\bibinfo {volume} {95}},\ \bibinfo
  {pages} {034001} (\bibinfo {year} {2017})}\BibitemShut {NoStop}%
\bibitem [{\citenamefont {Gras}\ \emph {et~al.}(2017)\citenamefont {Gras},
  \citenamefont {H{\"o}che}, \citenamefont {Kar}, \citenamefont {Larkoski},
  \citenamefont {L{\"o}nnblad}, \citenamefont {Pl{\"a}tzer}, \citenamefont
  {Si{\'o}dmok}, \citenamefont {Skands}, \citenamefont {Soyez},\ and\
  \citenamefont {Thaler}}]{Gras:2017ul}%
  \BibitemOpen
  \bibfield  {author} {\bibinfo {author} {\bibfnamefont {P.}~\bibnamefont
  {Gras}}, \bibinfo {author} {\bibfnamefont {S.}~\bibnamefont {H{\"o}che}},
  \bibinfo {author} {\bibfnamefont {D.}~\bibnamefont {Kar}}, \bibinfo {author}
  {\bibfnamefont {A.}~\bibnamefont {Larkoski}}, \bibinfo {author}
  {\bibfnamefont {L.}~\bibnamefont {L{\"o}nnblad}}, \bibinfo {author}
  {\bibfnamefont {S.}~\bibnamefont {Pl{\"a}tzer}}, \bibinfo {author}
  {\bibfnamefont {A.}~\bibnamefont {Si{\'o}dmok}}, \bibinfo {author}
  {\bibfnamefont {P.}~\bibnamefont {Skands}}, \bibinfo {author} {\bibfnamefont
  {G.}~\bibnamefont {Soyez}}, \ and\ \bibinfo {author} {\bibfnamefont
  {J.}~\bibnamefont {Thaler}},\ }\href {\doibase 10.1007/JHEP07(2017)091}
  {\bibfield  {journal} {\bibinfo  {journal} {Journal of High Energy Physics}\
  }\textbf {\bibinfo {volume} {2017}},\ \bibinfo {pages} {91} (\bibinfo {year}
  {2017})}\BibitemShut {NoStop}%
\bibitem [{\citenamefont {Kaplan}\ \emph {et~al.}(2008)\citenamefont {Kaplan},
  \citenamefont {Rehermann}, \citenamefont {Schwartz},\ and\ \citenamefont
  {Tweedie}}]{PhysRevLett.101.142001}%
  \BibitemOpen
  \bibfield  {author} {\bibinfo {author} {\bibfnamefont {D.~E.}\ \bibnamefont
  {Kaplan}}, \bibinfo {author} {\bibfnamefont {K.}~\bibnamefont {Rehermann}},
  \bibinfo {author} {\bibfnamefont {M.~D.}\ \bibnamefont {Schwartz}}, \ and\
  \bibinfo {author} {\bibfnamefont {B.}~\bibnamefont {Tweedie}},\ }\href
  {\doibase 10.1103/PhysRevLett.101.142001} {\bibfield  {journal} {\bibinfo
  {journal} {Phys. Rev. Lett.}\ }\textbf {\bibinfo {volume} {101}},\ \bibinfo
  {pages} {142001} (\bibinfo {year} {2008})}\BibitemShut {NoStop}%
\bibitem [{\citenamefont {Plehn}\ \emph {et~al.}(2012)\citenamefont {Plehn},
  \citenamefont {Spannowsky},\ and\ \citenamefont
  {Takeuchi}}]{PhysRevD.85.034029}%
  \BibitemOpen
  \bibfield  {author} {\bibinfo {author} {\bibfnamefont {T.}~\bibnamefont
  {Plehn}}, \bibinfo {author} {\bibfnamefont {M.}~\bibnamefont {Spannowsky}}, \
  and\ \bibinfo {author} {\bibfnamefont {M.}~\bibnamefont {Takeuchi}},\ }\href
  {\doibase 10.1103/PhysRevD.85.034029} {\bibfield  {journal} {\bibinfo
  {journal} {Phys. Rev. D}\ }\textbf {\bibinfo {volume} {85}},\ \bibinfo
  {pages} {034029} (\bibinfo {year} {2012})}\BibitemShut {NoStop}%
\bibitem [{\citenamefont {Soper}\ and\ \citenamefont
  {Spannowsky}(2013)}]{PhysRevD.87.054012}%
  \BibitemOpen
  \bibfield  {author} {\bibinfo {author} {\bibfnamefont {D.~E.}\ \bibnamefont
  {Soper}}\ and\ \bibinfo {author} {\bibfnamefont {M.}~\bibnamefont
  {Spannowsky}},\ }\href {\doibase 10.1103/PhysRevD.87.054012} {\bibfield
  {journal} {\bibinfo  {journal} {Phys. Rev. D}\ }\textbf {\bibinfo {volume}
  {87}},\ \bibinfo {pages} {054012} (\bibinfo {year} {2013})}\BibitemShut
  {NoStop}%
\bibitem [{\citenamefont {Anders}\ \emph {et~al.}(2014)\citenamefont {Anders},
  \citenamefont {Bernaciak}, \citenamefont {Kasieczka}, \citenamefont {Plehn},\
  and\ \citenamefont {Schell}}]{PhysRevD.89.074047}%
  \BibitemOpen
  \bibfield  {author} {\bibinfo {author} {\bibfnamefont {C.}~\bibnamefont
  {Anders}}, \bibinfo {author} {\bibfnamefont {C.}~\bibnamefont {Bernaciak}},
  \bibinfo {author} {\bibfnamefont {G.}~\bibnamefont {Kasieczka}}, \bibinfo
  {author} {\bibfnamefont {T.}~\bibnamefont {Plehn}}, \ and\ \bibinfo {author}
  {\bibfnamefont {T.}~\bibnamefont {Schell}},\ }\href {\doibase
  10.1103/PhysRevD.89.074047} {\bibfield  {journal} {\bibinfo  {journal} {Phys.
  Rev. D}\ }\textbf {\bibinfo {volume} {89}},\ \bibinfo {pages} {074047}
  (\bibinfo {year} {2014})}\BibitemShut {NoStop}%
\bibitem [{\citenamefont {Kasieczka}\ \emph {et~al.}(2015)\citenamefont
  {Kasieczka}, \citenamefont {Plehn}, \citenamefont {Schell}, \citenamefont
  {Strebler},\ and\ \citenamefont {Salam}}]{Kasieczka:2015tn}%
  \BibitemOpen
  \bibfield  {author} {\bibinfo {author} {\bibfnamefont {G.}~\bibnamefont
  {Kasieczka}}, \bibinfo {author} {\bibfnamefont {T.}~\bibnamefont {Plehn}},
  \bibinfo {author} {\bibfnamefont {T.}~\bibnamefont {Schell}}, \bibinfo
  {author} {\bibfnamefont {T.}~\bibnamefont {Strebler}}, \ and\ \bibinfo
  {author} {\bibfnamefont {G.~P.}\ \bibnamefont {Salam}},\ }\href {\doibase
  10.1007/JHEP06(2015)203} {\bibfield  {journal} {\bibinfo  {journal} {Journal
  of High Energy Physics}\ }\textbf {\bibinfo {volume} {2015}},\ \bibinfo
  {pages} {203} (\bibinfo {year} {2015})}\BibitemShut {NoStop}%
\bibitem [{\citenamefont {Thaler}\ and\ \citenamefont
  {Van~Tilburg}(2012)}]{Thaler:2012tb}%
  \BibitemOpen
  \bibfield  {author} {\bibinfo {author} {\bibfnamefont {J.}~\bibnamefont
  {Thaler}}\ and\ \bibinfo {author} {\bibfnamefont {K.}~\bibnamefont
  {Van~Tilburg}},\ }\href {\doibase 10.1007/JHEP02(2012)093} {\bibfield
  {journal} {\bibinfo  {journal} {Journal of High Energy Physics}\ }\textbf
  {\bibinfo {volume} {2012}},\ \bibinfo {pages} {93} (\bibinfo {year}
  {2012})}\BibitemShut {NoStop}%
\bibitem [{\citenamefont {G{\'a}ndara}\ and\ \citenamefont
  {Collaboration}(2009)}]{Gandara:2009vr}%
  \BibitemOpen
  \bibfield  {author} {\bibinfo {author} {\bibfnamefont {M.~G.}\ \bibnamefont
  {G{\'a}ndara}}\ and\ \bibinfo {author} {\bibfnamefont {t.~L.}\ \bibnamefont
  {Collaboration}},\ }\bibfield  {booktitle} {\emph {\bibinfo {booktitle}
  {Journal of Physics: Conference Series}},\ }\href {\doibase
  10.1088/1742-6596/171/1/012103} {\ \textbf {\bibinfo {volume} {171}},\
  \bibinfo {pages} {012103} (\bibinfo {year} {2009})}\BibitemShut {NoStop}%
\bibitem [{ATL(2016)}]{ATLAS:2016wv}%
  \BibitemOpen
  \bibfield  {booktitle} {\emph {\bibinfo {booktitle} {Journal of
  Instrumentation}},\ }\href {\doibase 10.1088/1748-0221/11/04/p04008} {\
  \textbf {\bibinfo {volume} {11}},\ \bibinfo {pages} {P04008} (\bibinfo {year}
  {2016})}\BibitemShut {NoStop}%
\bibitem [{\citenamefont {Aad}\ and\ \citenamefont {et~al. {ATLAS
  Collaboration}}(2019)}]{Aad:2019we}%
  \BibitemOpen
  \bibfield  {author} {\bibinfo {author} {\bibfnamefont {G.}~\bibnamefont
  {Aad}}\ and\ \bibinfo {author} {\bibnamefont {et~al. {ATLAS
  Collaboration}}},\ }\href {\doibase 10.1140/epjc/s10052-019-7450-8}
  {\bibfield  {journal} {\bibinfo  {journal} {The European Physical Journal C}\
  }\textbf {\bibinfo {volume} {79}},\ \bibinfo {pages} {970} (\bibinfo {year}
  {2019})}\BibitemShut {NoStop}%
\bibitem [{\citenamefont {de~Oliveira}\ \emph {et~al.}(2016)\citenamefont
  {de~Oliveira}, \citenamefont {Kagan}, \citenamefont {Mackey}, \citenamefont
  {Nachman},\ and\ \citenamefont {Schwartzman}}]{Oliveira:2016vm}%
  \BibitemOpen
  \bibfield  {author} {\bibinfo {author} {\bibfnamefont {L.}~\bibnamefont
  {de~Oliveira}}, \bibinfo {author} {\bibfnamefont {M.}~\bibnamefont {Kagan}},
  \bibinfo {author} {\bibfnamefont {L.}~\bibnamefont {Mackey}}, \bibinfo
  {author} {\bibfnamefont {B.}~\bibnamefont {Nachman}}, \ and\ \bibinfo
  {author} {\bibfnamefont {A.}~\bibnamefont {Schwartzman}},\ }\href {\doibase
  10.1007/JHEP07(2016)069} {\bibfield  {journal} {\bibinfo  {journal} {Journal
  of High Energy Physics}\ }\textbf {\bibinfo {volume} {2016}},\ \bibinfo
  {pages} {69} (\bibinfo {year} {2016})}\BibitemShut {NoStop}%
\bibitem [{\citenamefont {Komiske}\ \emph {et~al.}(2017)\citenamefont
  {Komiske}, \citenamefont {Metodiev},\ and\ \citenamefont
  {Schwartz}}]{Komiske:2017wc}%
  \BibitemOpen
  \bibfield  {author} {\bibinfo {author} {\bibfnamefont {P.~T.}\ \bibnamefont
  {Komiske}}, \bibinfo {author} {\bibfnamefont {E.~M.}\ \bibnamefont
  {Metodiev}}, \ and\ \bibinfo {author} {\bibfnamefont {M.~D.}\ \bibnamefont
  {Schwartz}},\ }\href {\doibase 10.1007/JHEP01(2017)110} {\bibfield  {journal}
  {\bibinfo  {journal} {Journal of High Energy Physics}\ }\textbf {\bibinfo
  {volume} {2017}},\ \bibinfo {pages} {110} (\bibinfo {year}
  {2017})}\BibitemShut {NoStop}%
\bibitem [{\citenamefont {Macaluso}\ and\ \citenamefont
  {Shih}(2018)}]{Macaluso:2018to}%
  \BibitemOpen
  \bibfield  {author} {\bibinfo {author} {\bibfnamefont {S.}~\bibnamefont
  {Macaluso}}\ and\ \bibinfo {author} {\bibfnamefont {D.}~\bibnamefont
  {Shih}},\ }\href {\doibase 10.1007/JHEP10(2018)121} {\bibfield  {journal}
  {\bibinfo  {journal} {Journal of High Energy Physics}\ }\textbf {\bibinfo
  {volume} {2018}},\ \bibinfo {pages} {121} (\bibinfo {year}
  {2018})}\BibitemShut {NoStop}%
\bibitem [{\citenamefont {Kasieczka}\ \emph {et~al.}(2017)\citenamefont
  {Kasieczka}, \citenamefont {Plehn}, \citenamefont {Russell},\ and\
  \citenamefont {Schell}}]{Kasieczka:2017vt}%
  \BibitemOpen
  \bibfield  {author} {\bibinfo {author} {\bibfnamefont {G.}~\bibnamefont
  {Kasieczka}}, \bibinfo {author} {\bibfnamefont {T.}~\bibnamefont {Plehn}},
  \bibinfo {author} {\bibfnamefont {M.}~\bibnamefont {Russell}}, \ and\
  \bibinfo {author} {\bibfnamefont {T.}~\bibnamefont {Schell}},\ }\href
  {\doibase 10.1007/JHEP05(2017)006} {\bibfield  {journal} {\bibinfo  {journal}
  {Journal of High Energy Physics}\ }\textbf {\bibinfo {volume} {2017}},\
  \bibinfo {pages} {6} (\bibinfo {year} {2017})}\BibitemShut {NoStop}%
\bibitem [{\citenamefont {Schwartzman}\ \emph {et~al.}(2016)\citenamefont
  {Schwartzman}, \citenamefont {Kagan}, \citenamefont {Mackey}, \citenamefont
  {Nachman},\ and\ \citenamefont {De~Oliveira}}]{Schwartzman:2016tu}%
  \BibitemOpen
  \bibfield  {author} {\bibinfo {author} {\bibfnamefont {A.}~\bibnamefont
  {Schwartzman}}, \bibinfo {author} {\bibfnamefont {M.}~\bibnamefont {Kagan}},
  \bibinfo {author} {\bibfnamefont {L.}~\bibnamefont {Mackey}}, \bibinfo
  {author} {\bibfnamefont {B.}~\bibnamefont {Nachman}}, \ and\ \bibinfo
  {author} {\bibfnamefont {L.}~\bibnamefont {De~Oliveira}},\ }\bibfield
  {booktitle} {\emph {\bibinfo {booktitle} {Journal of Physics: Conference
  Series}},\ }\href {\doibase 10.1088/1742-6596/762/1/012035} {\ \textbf
  {\bibinfo {volume} {762}},\ \bibinfo {pages} {012035} (\bibinfo {year}
  {2016})}\BibitemShut {NoStop}%
\bibitem [{\citenamefont {Louppe}\ \emph {et~al.}(2019)\citenamefont {Louppe},
  \citenamefont {Cho}, \citenamefont {Becot},\ and\ \citenamefont
  {Cranmer}}]{Louppe:2019vz}%
  \BibitemOpen
  \bibfield  {author} {\bibinfo {author} {\bibfnamefont {G.}~\bibnamefont
  {Louppe}}, \bibinfo {author} {\bibfnamefont {K.}~\bibnamefont {Cho}},
  \bibinfo {author} {\bibfnamefont {C.}~\bibnamefont {Becot}}, \ and\ \bibinfo
  {author} {\bibfnamefont {K.}~\bibnamefont {Cranmer}},\ }\href {\doibase
  10.1007/JHEP01(2019)057} {\bibfield  {journal} {\bibinfo  {journal} {Journal
  of High Energy Physics}\ }\textbf {\bibinfo {volume} {2019}},\ \bibinfo
  {pages} {57} (\bibinfo {year} {2019})}\BibitemShut {NoStop}%
\bibitem [{\citenamefont {Egan}\ \emph {et~al.}(2017)\citenamefont {Egan},
  \citenamefont {Fedorko}, \citenamefont {Lister}, \citenamefont {Pearkes},\
  and\ \citenamefont {Gay}}]{egan2017long}%
  \BibitemOpen
  \bibfield  {author} {\bibinfo {author} {\bibfnamefont {S.}~\bibnamefont
  {Egan}}, \bibinfo {author} {\bibfnamefont {W.}~\bibnamefont {Fedorko}},
  \bibinfo {author} {\bibfnamefont {A.}~\bibnamefont {Lister}}, \bibinfo
  {author} {\bibfnamefont {J.}~\bibnamefont {Pearkes}}, \ and\ \bibinfo
  {author} {\bibfnamefont {C.}~\bibnamefont {Gay}},\ }\href@noop {} {\bibfield
  {journal} {\bibinfo  {journal} {arXiv preprint arXiv:1711.09059}\ } (\bibinfo
  {year} {2017})}\BibitemShut {NoStop}%
\bibitem [{\citenamefont {Fraser}\ and\ \citenamefont
  {Schwartz}(2018)}]{Fraser:2018uo}%
  \BibitemOpen
  \bibfield  {author} {\bibinfo {author} {\bibfnamefont {K.}~\bibnamefont
  {Fraser}}\ and\ \bibinfo {author} {\bibfnamefont {M.~D.}\ \bibnamefont
  {Schwartz}},\ }\href {\doibase 10.1007/JHEP10(2018)093} {\bibfield  {journal}
  {\bibinfo  {journal} {Journal of High Energy Physics}\ }\textbf {\bibinfo
  {volume} {2018}},\ \bibinfo {pages} {93} (\bibinfo {year}
  {2018})}\BibitemShut {NoStop}%
\bibitem [{\citenamefont {Cheng}(2018)}]{Cheng:2018tr}%
  \BibitemOpen
  \bibfield  {author} {\bibinfo {author} {\bibfnamefont {T.}~\bibnamefont
  {Cheng}},\ }\href {\doibase 10.1007/s41781-018-0007-y} {\bibfield  {journal}
  {\bibinfo  {journal} {Computing and Software for Big Science}\ }\textbf
  {\bibinfo {volume} {2}},\ \bibinfo {pages} {3} (\bibinfo {year}
  {2018})}\BibitemShut {NoStop}%
\bibitem [{\citenamefont {Baldi}\ \emph {et~al.}(2016)\citenamefont {Baldi},
  \citenamefont {Bauer}, \citenamefont {Eng}, \citenamefont {Sadowski},\ and\
  \citenamefont {Whiteson}}]{PhysRevD.93.094034}%
  \BibitemOpen
  \bibfield  {author} {\bibinfo {author} {\bibfnamefont {P.}~\bibnamefont
  {Baldi}}, \bibinfo {author} {\bibfnamefont {K.}~\bibnamefont {Bauer}},
  \bibinfo {author} {\bibfnamefont {C.}~\bibnamefont {Eng}}, \bibinfo {author}
  {\bibfnamefont {P.}~\bibnamefont {Sadowski}}, \ and\ \bibinfo {author}
  {\bibfnamefont {D.}~\bibnamefont {Whiteson}},\ }\href {\doibase
  10.1103/PhysRevD.93.094034} {\bibfield  {journal} {\bibinfo  {journal} {Phys.
  Rev. D}\ }\textbf {\bibinfo {volume} {93}},\ \bibinfo {pages} {094034}
  (\bibinfo {year} {2016})}\BibitemShut {NoStop}%
\bibitem [{\citenamefont {Barnard}\ \emph {et~al.}(2017)\citenamefont
  {Barnard}, \citenamefont {Dawe}, \citenamefont {Dolan},\ and\ \citenamefont
  {Rajcic}}]{PhysRevD.95.014018}%
  \BibitemOpen
  \bibfield  {author} {\bibinfo {author} {\bibfnamefont {J.}~\bibnamefont
  {Barnard}}, \bibinfo {author} {\bibfnamefont {E.~N.}\ \bibnamefont {Dawe}},
  \bibinfo {author} {\bibfnamefont {M.~J.}\ \bibnamefont {Dolan}}, \ and\
  \bibinfo {author} {\bibfnamefont {N.}~\bibnamefont {Rajcic}},\ }\href
  {\doibase 10.1103/PhysRevD.95.014018} {\bibfield  {journal} {\bibinfo
  {journal} {Phys. Rev. D}\ }\textbf {\bibinfo {volume} {95}},\ \bibinfo
  {pages} {014018} (\bibinfo {year} {2017})}\BibitemShut {NoStop}%
\bibitem [{\citenamefont {Guest}\ \emph {et~al.}(2016)\citenamefont {Guest},
  \citenamefont {Collado}, \citenamefont {Baldi}, \citenamefont {Hsu},
  \citenamefont {Urban},\ and\ \citenamefont {Whiteson}}]{PhysRevD.94.112002}%
  \BibitemOpen
  \bibfield  {author} {\bibinfo {author} {\bibfnamefont {D.}~\bibnamefont
  {Guest}}, \bibinfo {author} {\bibfnamefont {J.}~\bibnamefont {Collado}},
  \bibinfo {author} {\bibfnamefont {P.}~\bibnamefont {Baldi}}, \bibinfo
  {author} {\bibfnamefont {S.-C.}\ \bibnamefont {Hsu}}, \bibinfo {author}
  {\bibfnamefont {G.}~\bibnamefont {Urban}}, \ and\ \bibinfo {author}
  {\bibfnamefont {D.}~\bibnamefont {Whiteson}},\ }\href {\doibase
  10.1103/PhysRevD.94.112002} {\bibfield  {journal} {\bibinfo  {journal} {Phys.
  Rev. D}\ }\textbf {\bibinfo {volume} {94}},\ \bibinfo {pages} {112002}
  (\bibinfo {year} {2016})}\BibitemShut {NoStop}%
\bibitem [{\citenamefont {Komiske}\ \emph {et~al.}(2019)\citenamefont
  {Komiske}, \citenamefont {Metodiev},\ and\ \citenamefont
  {Thaler}}]{Komiske:2019um}%
  \BibitemOpen
  \bibfield  {author} {\bibinfo {author} {\bibfnamefont {P.~T.}\ \bibnamefont
  {Komiske}}, \bibinfo {author} {\bibfnamefont {E.~M.}\ \bibnamefont
  {Metodiev}}, \ and\ \bibinfo {author} {\bibfnamefont {J.}~\bibnamefont
  {Thaler}},\ }\href {\doibase 10.1007/JHEP01(2019)121} {\bibfield  {journal}
  {\bibinfo  {journal} {Journal of High Energy Physics}\ }\textbf {\bibinfo
  {volume} {2019}},\ \bibinfo {pages} {121} (\bibinfo {year}
  {2019})}\BibitemShut {NoStop}%
\bibitem [{\citenamefont {Dolan}\ and\ \citenamefont
  {Ore}(2021)}]{PhysRevD.103.074022}%
  \BibitemOpen
  \bibfield  {author} {\bibinfo {author} {\bibfnamefont {M.~J.}\ \bibnamefont
  {Dolan}}\ and\ \bibinfo {author} {\bibfnamefont {A.}~\bibnamefont {Ore}},\
  }\href {\doibase 10.1103/PhysRevD.103.074022} {\bibfield  {journal} {\bibinfo
   {journal} {Phys. Rev. D}\ }\textbf {\bibinfo {volume} {103}},\ \bibinfo
  {pages} {074022} (\bibinfo {year} {2021})}\BibitemShut {NoStop}%
\bibitem [{\citenamefont {Moreno}\ \emph {et~al.}(2020)\citenamefont {Moreno},
  \citenamefont {Cerri}, \citenamefont {Duarte}, \citenamefont {Newman},
  \citenamefont {Nguyen}, \citenamefont {Periwal}, \citenamefont {Pierini},
  \citenamefont {Serikova}, \citenamefont {Spiropulu},\ and\ \citenamefont
  {Vlimant}}]{Moreno:2020up}%
  \BibitemOpen
  \bibfield  {author} {\bibinfo {author} {\bibfnamefont {E.~A.}\ \bibnamefont
  {Moreno}}, \bibinfo {author} {\bibfnamefont {O.}~\bibnamefont {Cerri}},
  \bibinfo {author} {\bibfnamefont {J.~M.}\ \bibnamefont {Duarte}}, \bibinfo
  {author} {\bibfnamefont {H.~B.}\ \bibnamefont {Newman}}, \bibinfo {author}
  {\bibfnamefont {T.~Q.}\ \bibnamefont {Nguyen}}, \bibinfo {author}
  {\bibfnamefont {A.}~\bibnamefont {Periwal}}, \bibinfo {author} {\bibfnamefont
  {M.}~\bibnamefont {Pierini}}, \bibinfo {author} {\bibfnamefont
  {A.}~\bibnamefont {Serikova}}, \bibinfo {author} {\bibfnamefont
  {M.}~\bibnamefont {Spiropulu}}, \ and\ \bibinfo {author} {\bibfnamefont
  {J.-R.}\ \bibnamefont {Vlimant}},\ }\href {\doibase
  10.1140/epjc/s10052-020-7608-4} {\bibfield  {journal} {\bibinfo  {journal}
  {The European Physical Journal C}\ }\textbf {\bibinfo {volume} {80}},\
  \bibinfo {pages} {58} (\bibinfo {year} {2020})}\BibitemShut {NoStop}%
\bibitem [{\citenamefont {Farrell}\ \emph {et~al.}(2018)\citenamefont
  {Farrell}, \citenamefont {Calafiura}, \citenamefont {Mudigonda},
  \citenamefont {{Prabhat}}, \citenamefont {Anderson}, \citenamefont {Vlimant},
  \citenamefont {Zheng}, \citenamefont {Bendavid}, \citenamefont {Spiropulu},
  \citenamefont {Cerati}, \citenamefont {Gray}, \citenamefont {Kowalkowski},
  \citenamefont {Spentzouris},\ and\ \citenamefont {Tsaris}}]{2018Novel}%
  \BibitemOpen
  \bibfield  {author} {\bibinfo {author} {\bibfnamefont {S.}~\bibnamefont
  {Farrell}}, \bibinfo {author} {\bibfnamefont {P.}~\bibnamefont {Calafiura}},
  \bibinfo {author} {\bibfnamefont {M.}~\bibnamefont {Mudigonda}}, \bibinfo
  {author} {\bibnamefont {{Prabhat}}}, \bibinfo {author} {\bibfnamefont
  {D.}~\bibnamefont {Anderson}}, \bibinfo {author} {\bibfnamefont {J.-R.}\
  \bibnamefont {Vlimant}}, \bibinfo {author} {\bibfnamefont {S.}~\bibnamefont
  {Zheng}}, \bibinfo {author} {\bibfnamefont {J.}~\bibnamefont {Bendavid}},
  \bibinfo {author} {\bibfnamefont {M.}~\bibnamefont {Spiropulu}}, \bibinfo
  {author} {\bibfnamefont {G.}~\bibnamefont {Cerati}}, \bibinfo {author}
  {\bibfnamefont {L.}~\bibnamefont {Gray}}, \bibinfo {author} {\bibfnamefont
  {J.}~\bibnamefont {Kowalkowski}}, \bibinfo {author} {\bibfnamefont
  {P.}~\bibnamefont {Spentzouris}}, \ and\ \bibinfo {author} {\bibfnamefont
  {A.}~\bibnamefont {Tsaris}},\ }\href {\doibase 10.48550/ARXIV.1810.06111} {\
  (\bibinfo {year} {2018}),\ 10.48550/ARXIV.1810.06111}\BibitemShut {NoStop}%
\bibitem [{\citenamefont {Arjona~Mart{\'\i}nez}\ \emph
  {et~al.}(2019)\citenamefont {Arjona~Mart{\'\i}nez}, \citenamefont {Cerri},
  \citenamefont {Spiropulu}, \citenamefont {Vlimant},\ and\ \citenamefont
  {Pierini}}]{Arjona-Martinez:2019vx}%
  \BibitemOpen
  \bibfield  {author} {\bibinfo {author} {\bibfnamefont {J.}~\bibnamefont
  {Arjona~Mart{\'\i}nez}}, \bibinfo {author} {\bibfnamefont {O.}~\bibnamefont
  {Cerri}}, \bibinfo {author} {\bibfnamefont {M.}~\bibnamefont {Spiropulu}},
  \bibinfo {author} {\bibfnamefont {J.~R.}\ \bibnamefont {Vlimant}}, \ and\
  \bibinfo {author} {\bibfnamefont {M.}~\bibnamefont {Pierini}},\ }\href
  {\doibase 10.1140/epjp/i2019-12710-3} {\bibfield  {journal} {\bibinfo
  {journal} {The European Physical Journal Plus}\ }\textbf {\bibinfo {volume}
  {134}},\ \bibinfo {pages} {333} (\bibinfo {year} {2019})}\BibitemShut
  {NoStop}%
\bibitem [{\citenamefont {Qasim}\ \emph {et~al.}(2019)\citenamefont {Qasim},
  \citenamefont {Kieseler}, \citenamefont {Iiyama},\ and\ \citenamefont
  {Pierini}}]{Qasim:2019ux}%
  \BibitemOpen
  \bibfield  {author} {\bibinfo {author} {\bibfnamefont {S.~R.}\ \bibnamefont
  {Qasim}}, \bibinfo {author} {\bibfnamefont {J.}~\bibnamefont {Kieseler}},
  \bibinfo {author} {\bibfnamefont {Y.}~\bibnamefont {Iiyama}}, \ and\ \bibinfo
  {author} {\bibfnamefont {M.}~\bibnamefont {Pierini}},\ }\href {\doibase
  10.1140/epjc/s10052-019-7113-9} {\bibfield  {journal} {\bibinfo  {journal}
  {The European Physical Journal C}\ }\textbf {\bibinfo {volume} {79}},\
  \bibinfo {pages} {608} (\bibinfo {year} {2019})}\BibitemShut {NoStop}%
\bibitem [{\citenamefont {Abdughani}\ \emph {et~al.}(2019)\citenamefont
  {Abdughani}, \citenamefont {Ren}, \citenamefont {Wu},\ and\ \citenamefont
  {Yang}}]{Abdughani:2019uk}%
  \BibitemOpen
  \bibfield  {author} {\bibinfo {author} {\bibfnamefont {M.}~\bibnamefont
  {Abdughani}}, \bibinfo {author} {\bibfnamefont {J.}~\bibnamefont {Ren}},
  \bibinfo {author} {\bibfnamefont {L.}~\bibnamefont {Wu}}, \ and\ \bibinfo
  {author} {\bibfnamefont {J.~M.}\ \bibnamefont {Yang}},\ }\href {\doibase
  10.1007/JHEP08(2019)055} {\bibfield  {journal} {\bibinfo  {journal} {Journal
  of High Energy Physics}\ }\textbf {\bibinfo {volume} {2019}},\ \bibinfo
  {pages} {55} (\bibinfo {year} {2019})}\BibitemShut {NoStop}%
\bibitem [{\citenamefont {Ren}\ \emph {et~al.}(2020)\citenamefont {Ren},
  \citenamefont {Wu},\ and\ \citenamefont {Yang}}]{Ren:2020tu}%
  \BibitemOpen
  \bibfield  {author} {\bibinfo {author} {\bibfnamefont {J.}~\bibnamefont
  {Ren}}, \bibinfo {author} {\bibfnamefont {L.}~\bibnamefont {Wu}}, \ and\
  \bibinfo {author} {\bibfnamefont {J.~M.}\ \bibnamefont {Yang}},\ }\href
  {\doibase https://doi.org/10.1016/j.physletb.2020.135198} {\bibfield
  {journal} {\bibinfo  {journal} {Physics Letters B}\ }\textbf {\bibinfo
  {volume} {802}},\ \bibinfo {pages} {135198} (\bibinfo {year}
  {2020})}\BibitemShut {NoStop}%
\bibitem [{\citenamefont {Henrion}\ \emph {et~al.}(2017)\citenamefont
  {Henrion}, \citenamefont {Brehmer}, \citenamefont {Bruna}, \citenamefont
  {Cho}, \citenamefont {Cranmer}, \citenamefont {Louppe},\ and\ \citenamefont
  {Rochette}}]{henrion2017neural}%
  \BibitemOpen
  \bibfield  {author} {\bibinfo {author} {\bibfnamefont {I.}~\bibnamefont
  {Henrion}}, \bibinfo {author} {\bibfnamefont {J.}~\bibnamefont {Brehmer}},
  \bibinfo {author} {\bibfnamefont {J.}~\bibnamefont {Bruna}}, \bibinfo
  {author} {\bibfnamefont {K.}~\bibnamefont {Cho}}, \bibinfo {author}
  {\bibfnamefont {K.}~\bibnamefont {Cranmer}}, \bibinfo {author} {\bibfnamefont
  {G.}~\bibnamefont {Louppe}}, \ and\ \bibinfo {author} {\bibfnamefont
  {G.}~\bibnamefont {Rochette}},\ }\href@noop {} {\  (\bibinfo {year}
  {2017})}\BibitemShut {NoStop}%
\bibitem [{\citenamefont {Qu}\ and\ \citenamefont {Gouskos}(2020)}]{qu2020jet}%
  \BibitemOpen
  \bibfield  {author} {\bibinfo {author} {\bibfnamefont {H.}~\bibnamefont
  {Qu}}\ and\ \bibinfo {author} {\bibfnamefont {L.}~\bibnamefont {Gouskos}},\
  }\href {\doibase 10.1103/PhysRevD.101.056019} {\bibfield  {journal} {\bibinfo
   {journal} {Phys. Rev. D}\ }\textbf {\bibinfo {volume} {101}},\ \bibinfo
  {pages} {056019} (\bibinfo {year} {2020})}\BibitemShut {NoStop}%
\bibitem [{\citenamefont {Atkinson}\ \emph {et~al.}(2021)\citenamefont
  {Atkinson}, \citenamefont {Bhardwaj}, \citenamefont {Englert}, \citenamefont
  {Ngairangbam},\ and\ \citenamefont {Spannowsky}}]{Atkinson:2021tp}%
  \BibitemOpen
  \bibfield  {author} {\bibinfo {author} {\bibfnamefont {O.}~\bibnamefont
  {Atkinson}}, \bibinfo {author} {\bibfnamefont {A.}~\bibnamefont {Bhardwaj}},
  \bibinfo {author} {\bibfnamefont {C.}~\bibnamefont {Englert}}, \bibinfo
  {author} {\bibfnamefont {V.~S.}\ \bibnamefont {Ngairangbam}}, \ and\ \bibinfo
  {author} {\bibfnamefont {M.}~\bibnamefont {Spannowsky}},\ }\href {\doibase
  10.1007/JHEP08(2021)080} {\bibfield  {journal} {\bibinfo  {journal} {Journal
  of High Energy Physics}\ }\textbf {\bibinfo {volume} {2021}},\ \bibinfo
  {pages} {80} (\bibinfo {year} {2021})}\BibitemShut {NoStop}%
\bibitem [{\citenamefont {Atkinson}\ \emph {et~al.}(2022)\citenamefont
  {Atkinson}, \citenamefont {Bhardwaj}, \citenamefont {Englert}, \citenamefont
  {Konar}, \citenamefont {Ngairangbam},\ and\ \citenamefont
  {Spannowsky}}]{10.3389/frai.2022.943135}%
  \BibitemOpen
  \bibfield  {author} {\bibinfo {author} {\bibfnamefont {O.}~\bibnamefont
  {Atkinson}}, \bibinfo {author} {\bibfnamefont {A.}~\bibnamefont {Bhardwaj}},
  \bibinfo {author} {\bibfnamefont {C.}~\bibnamefont {Englert}}, \bibinfo
  {author} {\bibfnamefont {P.}~\bibnamefont {Konar}}, \bibinfo {author}
  {\bibfnamefont {V.~S.}\ \bibnamefont {Ngairangbam}}, \ and\ \bibinfo {author}
  {\bibfnamefont {M.}~\bibnamefont {Spannowsky}},\ }\href {\doibase
  10.3389/frai.2022.943135} {\bibfield  {journal} {\bibinfo  {journal}
  {Frontiers in Artificial Intelligence}\ }\textbf {\bibinfo {volume} {5}}
  (\bibinfo {year} {2022}),\ 10.3389/frai.2022.943135}\BibitemShut {NoStop}%
\bibitem [{\citenamefont {Dreyer}\ and\ \citenamefont
  {Qu}(2021)}]{Dreyer:2021uo}%
  \BibitemOpen
  \bibfield  {author} {\bibinfo {author} {\bibfnamefont {F.~A.}\ \bibnamefont
  {Dreyer}}\ and\ \bibinfo {author} {\bibfnamefont {H.}~\bibnamefont {Qu}},\
  }\href {\doibase 10.1007/JHEP03(2021)052} {\bibfield  {journal} {\bibinfo
  {journal} {Journal of High Energy Physics}\ }\textbf {\bibinfo {volume}
  {2021}},\ \bibinfo {pages} {52} (\bibinfo {year} {2021})}\BibitemShut
  {NoStop}%
\bibitem [{\citenamefont {Dreyer}\ \emph
  {et~al.}(2022{\natexlab{a}})\citenamefont {Dreyer}, \citenamefont {Soyez},\
  and\ \citenamefont {Takacs}}]{Dreyer:2022ww}%
  \BibitemOpen
  \bibfield  {author} {\bibinfo {author} {\bibfnamefont {F.~A.}\ \bibnamefont
  {Dreyer}}, \bibinfo {author} {\bibfnamefont {G.}~\bibnamefont {Soyez}}, \
  and\ \bibinfo {author} {\bibfnamefont {A.}~\bibnamefont {Takacs}},\ }\href
  {\doibase 10.1007/JHEP08(2022)177} {\bibfield  {journal} {\bibinfo  {journal}
  {Journal of High Energy Physics}\ }\textbf {\bibinfo {volume} {2022}},\
  \bibinfo {pages} {177} (\bibinfo {year} {2022}{\natexlab{a}})}\BibitemShut
  {NoStop}%
\bibitem [{\citenamefont {Dreyer}\ \emph
  {et~al.}(2022{\natexlab{b}})\citenamefont {Dreyer}, \citenamefont
  {Grabarczyk},\ and\ \citenamefont {Monni}}]{Dreyer:2022wk}%
  \BibitemOpen
  \bibfield  {author} {\bibinfo {author} {\bibfnamefont {F.~A.}\ \bibnamefont
  {Dreyer}}, \bibinfo {author} {\bibfnamefont {R.}~\bibnamefont {Grabarczyk}},
  \ and\ \bibinfo {author} {\bibfnamefont {P.~F.}\ \bibnamefont {Monni}},\
  }\href {\doibase 10.1140/epjc/s10052-022-10469-9} {\bibfield  {journal}
  {\bibinfo  {journal} {The European Physical Journal C}\ }\textbf {\bibinfo
  {volume} {82}},\ \bibinfo {pages} {564} (\bibinfo {year}
  {2022}{\natexlab{b}})}\BibitemShut {NoStop}%
\bibitem [{\citenamefont {Dreyer}\ \emph {et~al.}(2018)\citenamefont {Dreyer},
  \citenamefont {Salam},\ and\ \citenamefont {Soyez}}]{Dreyer:2018ut}%
  \BibitemOpen
  \bibfield  {author} {\bibinfo {author} {\bibfnamefont {F.~A.}\ \bibnamefont
  {Dreyer}}, \bibinfo {author} {\bibfnamefont {G.~P.}\ \bibnamefont {Salam}}, \
  and\ \bibinfo {author} {\bibfnamefont {G.}~\bibnamefont {Soyez}},\ }\href
  {\doibase 10.1007/JHEP12(2018)064} {\bibfield  {journal} {\bibinfo  {journal}
  {Journal of High Energy Physics}\ }\textbf {\bibinfo {volume} {2018}},\
  \bibinfo {pages} {64} (\bibinfo {year} {2018})}\BibitemShut {NoStop}%
\bibitem [{\citenamefont {Li}\ \emph {et~al.}(2022)\citenamefont {Li},
  \citenamefont {Qu}, \citenamefont {Qian}, \citenamefont {Meng}, \citenamefont
  {Gong}, \citenamefont {Zhang}, \citenamefont {Liu},\ and\ \citenamefont
  {Li}}]{Li:2022xfc}%
  \BibitemOpen
  \bibfield  {author} {\bibinfo {author} {\bibfnamefont {C.}~\bibnamefont
  {Li}}, \bibinfo {author} {\bibfnamefont {H.}~\bibnamefont {Qu}}, \bibinfo
  {author} {\bibfnamefont {S.}~\bibnamefont {Qian}}, \bibinfo {author}
  {\bibfnamefont {Q.}~\bibnamefont {Meng}}, \bibinfo {author} {\bibfnamefont
  {S.}~\bibnamefont {Gong}}, \bibinfo {author} {\bibfnamefont {J.}~\bibnamefont
  {Zhang}}, \bibinfo {author} {\bibfnamefont {T.-Y.}\ \bibnamefont {Liu}}, \
  and\ \bibinfo {author} {\bibfnamefont {Q.}~\bibnamefont {Li}},\ }\href@noop
  {} {\  (\bibinfo {year} {2022})},\ \Eprint {http://arxiv.org/abs/2208.07814}
  {arXiv:2208.07814 [hep-ph]} \BibitemShut {NoStop}%
\bibitem [{\citenamefont {Gong}\ \emph {et~al.}(2022)\citenamefont {Gong},
  \citenamefont {Meng}, \citenamefont {Zhang}, \citenamefont {Qu},
  \citenamefont {Li}, \citenamefont {Qian}, \citenamefont {Du}, \citenamefont
  {Ma},\ and\ \citenamefont {Liu}}]{Gong:2022wk}%
  \BibitemOpen
  \bibfield  {author} {\bibinfo {author} {\bibfnamefont {S.}~\bibnamefont
  {Gong}}, \bibinfo {author} {\bibfnamefont {Q.}~\bibnamefont {Meng}}, \bibinfo
  {author} {\bibfnamefont {J.}~\bibnamefont {Zhang}}, \bibinfo {author}
  {\bibfnamefont {H.}~\bibnamefont {Qu}}, \bibinfo {author} {\bibfnamefont
  {C.}~\bibnamefont {Li}}, \bibinfo {author} {\bibfnamefont {S.}~\bibnamefont
  {Qian}}, \bibinfo {author} {\bibfnamefont {W.}~\bibnamefont {Du}}, \bibinfo
  {author} {\bibfnamefont {Z.-M.}\ \bibnamefont {Ma}}, \ and\ \bibinfo {author}
  {\bibfnamefont {T.-Y.}\ \bibnamefont {Liu}},\ }\href {\doibase
  10.1007/JHEP07(2022)030} {\bibfield  {journal} {\bibinfo  {journal} {Journal
  of High Energy Physics}\ }\textbf {\bibinfo {volume} {2022}},\ \bibinfo
  {pages} {30} (\bibinfo {year} {2022})}\BibitemShut {NoStop}%
\bibitem [{\citenamefont {Duvenaud}\ \emph {et~al.}(2015)\citenamefont
  {Duvenaud}, \citenamefont {Maclaurin}, \citenamefont {Iparraguirre},
  \citenamefont {Bombarell}, \citenamefont {Hirzel}, \citenamefont
  {Aspuru-Guzik},\ and\ \citenamefont {Adams}}]{NIPS2015_f9be311e}%
  \BibitemOpen
  \bibfield  {author} {\bibinfo {author} {\bibfnamefont {D.~K.}\ \bibnamefont
  {Duvenaud}}, \bibinfo {author} {\bibfnamefont {D.}~\bibnamefont {Maclaurin}},
  \bibinfo {author} {\bibfnamefont {J.}~\bibnamefont {Iparraguirre}}, \bibinfo
  {author} {\bibfnamefont {R.}~\bibnamefont {Bombarell}}, \bibinfo {author}
  {\bibfnamefont {T.}~\bibnamefont {Hirzel}}, \bibinfo {author} {\bibfnamefont
  {A.}~\bibnamefont {Aspuru-Guzik}}, \ and\ \bibinfo {author} {\bibfnamefont
  {R.~P.}\ \bibnamefont {Adams}},\ }in\ \href
  {https://proceedings.neurips.cc/paper/2015/file/f9be311e65d81a9ad8150a60844bb94c-Paper.pdf}
  {\emph {\bibinfo {booktitle} {Advances in Neural Information Processing
  Systems}}},\ Vol.~\bibinfo {volume} {28},\ \bibinfo {editor} {edited by\
  \bibinfo {editor} {\bibfnamefont {C.}~\bibnamefont {Cortes}}, \bibinfo
  {editor} {\bibfnamefont {N.}~\bibnamefont {Lawrence}}, \bibinfo {editor}
  {\bibfnamefont {D.}~\bibnamefont {Lee}}, \bibinfo {editor} {\bibfnamefont
  {M.}~\bibnamefont {Sugiyama}}, \ and\ \bibinfo {editor} {\bibfnamefont
  {R.}~\bibnamefont {Garnett}}}\ (\bibinfo  {publisher} {Curran Associates,
  Inc.},\ \bibinfo {year} {2015})\BibitemShut {NoStop}%
\bibitem [{\citenamefont {Kushnir}\ \emph {et~al.}(2006)\citenamefont
  {Kushnir}, \citenamefont {Galun},\ and\ \citenamefont
  {Brandt}}]{Kushnir:2006un}%
  \BibitemOpen
  \bibfield  {author} {\bibinfo {author} {\bibfnamefont {D.}~\bibnamefont
  {Kushnir}}, \bibinfo {author} {\bibfnamefont {M.}~\bibnamefont {Galun}}, \
  and\ \bibinfo {author} {\bibfnamefont {A.}~\bibnamefont {Brandt}},\
  }\bibfield  {booktitle} {\emph {\bibinfo {booktitle} {Similarity-based
  Pattern Recognition}},\ }\href {\doibase
  https://doi.org/10.1016/j.patcog.2006.04.007} {\bibfield  {journal} {\bibinfo
   {journal} {Pattern Recognition}\ }\textbf {\bibinfo {volume} {39}},\
  \bibinfo {pages} {1876} (\bibinfo {year} {2006})}\BibitemShut {NoStop}%
\bibitem [{\citenamefont {Rhee}\ \emph {et~al.}(2018)\citenamefont {Rhee},
  \citenamefont {Seo},\ and\ \citenamefont {Kim}}]{ijcai2018p490}%
  \BibitemOpen
  \bibfield  {author} {\bibinfo {author} {\bibfnamefont {S.}~\bibnamefont
  {Rhee}}, \bibinfo {author} {\bibfnamefont {S.}~\bibnamefont {Seo}}, \ and\
  \bibinfo {author} {\bibfnamefont {S.}~\bibnamefont {Kim}},\ }in\ \href
  {\doibase 10.24963/ijcai.2018/490} {\emph {\bibinfo {booktitle} {Proceedings
  of the Twenty-Seventh International Joint Conference on Artificial
  Intelligence, {IJCAI-18}}}}\ (\bibinfo  {publisher} {International Joint
  Conferences on Artificial Intelligence Organization},\ \bibinfo {year}
  {2018})\ pp.\ \bibinfo {pages} {3527--3534}\BibitemShut {NoStop}%
\bibitem [{\citenamefont {Defferrard}\ \emph {et~al.}(2016)\citenamefont
  {Defferrard}, \citenamefont {Bresson},\ and\ \citenamefont
  {Vandergheynst}}]{NIPS2016_04df4d43}%
  \BibitemOpen
  \bibfield  {author} {\bibinfo {author} {\bibfnamefont {M.}~\bibnamefont
  {Defferrard}}, \bibinfo {author} {\bibfnamefont {X.}~\bibnamefont {Bresson}},
  \ and\ \bibinfo {author} {\bibfnamefont {P.}~\bibnamefont {Vandergheynst}},\
  }in\ \href
  {https://proceedings.neurips.cc/paper/2016/file/04df4d434d481c5bb723be1b6df1ee65-Paper.pdf}
  {\emph {\bibinfo {booktitle} {Advances in Neural Information Processing
  Systems}}},\ Vol.~\bibinfo {volume} {29},\ \bibinfo {editor} {edited by\
  \bibinfo {editor} {\bibfnamefont {D.}~\bibnamefont {Lee}}, \bibinfo {editor}
  {\bibfnamefont {M.}~\bibnamefont {Sugiyama}}, \bibinfo {editor}
  {\bibfnamefont {U.}~\bibnamefont {Luxburg}}, \bibinfo {editor} {\bibfnamefont
  {I.}~\bibnamefont {Guyon}}, \ and\ \bibinfo {editor} {\bibfnamefont
  {R.}~\bibnamefont {Garnett}}}\ (\bibinfo  {publisher} {Curran Associates,
  Inc.},\ \bibinfo {year} {2016})\BibitemShut {NoStop}%
\bibitem [{\citenamefont {Ying}\ \emph {et~al.}(2018)\citenamefont {Ying},
  \citenamefont {You}, \citenamefont {Morris}, \citenamefont {Ren},
  \citenamefont {Hamilton},\ and\ \citenamefont
  {Leskovec}}]{NEURIPS2018_e77dbaf6}%
  \BibitemOpen
  \bibfield  {author} {\bibinfo {author} {\bibfnamefont {Z.}~\bibnamefont
  {Ying}}, \bibinfo {author} {\bibfnamefont {J.}~\bibnamefont {You}}, \bibinfo
  {author} {\bibfnamefont {C.}~\bibnamefont {Morris}}, \bibinfo {author}
  {\bibfnamefont {X.}~\bibnamefont {Ren}}, \bibinfo {author} {\bibfnamefont
  {W.}~\bibnamefont {Hamilton}}, \ and\ \bibinfo {author} {\bibfnamefont
  {J.}~\bibnamefont {Leskovec}},\ }in\ \href
  {https://proceedings.neurips.cc/paper/2018/file/e77dbaf6759253c7c6d0efc5690369c7-Paper.pdf}
  {\emph {\bibinfo {booktitle} {Advances in Neural Information Processing
  Systems}}},\ Vol.~\bibinfo {volume} {31},\ \bibinfo {editor} {edited by\
  \bibinfo {editor} {\bibfnamefont {S.}~\bibnamefont {Bengio}}, \bibinfo
  {editor} {\bibfnamefont {H.}~\bibnamefont {Wallach}}, \bibinfo {editor}
  {\bibfnamefont {H.}~\bibnamefont {Larochelle}}, \bibinfo {editor}
  {\bibfnamefont {K.}~\bibnamefont {Grauman}}, \bibinfo {editor} {\bibfnamefont
  {N.}~\bibnamefont {Cesa-Bianchi}}, \ and\ \bibinfo {editor} {\bibfnamefont
  {R.}~\bibnamefont {Garnett}}}\ (\bibinfo  {publisher} {Curran Associates,
  Inc.},\ \bibinfo {year} {2018})\BibitemShut {NoStop}%
\bibitem [{\citenamefont {Cangea}\ \emph {et~al.}(2018)\citenamefont {Cangea},
  \citenamefont {Veli{\v{c}}kovi{\'c}}, \citenamefont {Jovanovi{\'c}},
  \citenamefont {Kipf},\ and\ \citenamefont {Li{\`o}}}]{cangea2018towards}%
  \BibitemOpen
  \bibfield  {author} {\bibinfo {author} {\bibfnamefont {C.}~\bibnamefont
  {Cangea}}, \bibinfo {author} {\bibfnamefont {P.}~\bibnamefont
  {Veli{\v{c}}kovi{\'c}}}, \bibinfo {author} {\bibfnamefont {N.}~\bibnamefont
  {Jovanovi{\'c}}}, \bibinfo {author} {\bibfnamefont {T.}~\bibnamefont {Kipf}},
  \ and\ \bibinfo {author} {\bibfnamefont {P.}~\bibnamefont {Li{\`o}}},\
  }\href@noop {} {\bibfield  {journal} {\bibinfo  {journal} {arXiv preprint
  arXiv:1811.01287}\ } (\bibinfo {year} {2018})}\BibitemShut {NoStop}%
\bibitem [{\citenamefont {Noutahi}\ \emph {et~al.}(2019)\citenamefont
  {Noutahi}, \citenamefont {Beaini}, \citenamefont {Horwood}, \citenamefont
  {Gigu{\`e}re},\ and\ \citenamefont {Tossou}}]{noutahi2019t}%
  \BibitemOpen
  \bibfield  {author} {\bibinfo {author} {\bibfnamefont {E.}~\bibnamefont
  {Noutahi}}, \bibinfo {author} {\bibfnamefont {D.}~\bibnamefont {Beaini}},
  \bibinfo {author} {\bibfnamefont {J.}~\bibnamefont {Horwood}}, \bibinfo
  {author} {\bibfnamefont {S.}~\bibnamefont {Gigu{\`e}re}}, \ and\ \bibinfo
  {author} {\bibfnamefont {P.}~\bibnamefont {Tossou}},\ }\href {\doibase
  10.48550/ARXIV.1905.11577} {\enquote {\bibinfo {title} {Towards interpretable
  sparse graph representation learning with laplacian pooling},}\ } (\bibinfo
  {year} {2019})\BibitemShut {NoStop}%
\bibitem [{\citenamefont {Ma}\ \emph {et~al.}(2019)\citenamefont {Ma},
  \citenamefont {Wang}, \citenamefont {Aggarwal},\ and\ \citenamefont
  {Tang}}]{ma2019graph}%
  \BibitemOpen
  \bibfield  {author} {\bibinfo {author} {\bibfnamefont {Y.}~\bibnamefont
  {Ma}}, \bibinfo {author} {\bibfnamefont {S.}~\bibnamefont {Wang}}, \bibinfo
  {author} {\bibfnamefont {C.~C.}\ \bibnamefont {Aggarwal}}, \ and\ \bibinfo
  {author} {\bibfnamefont {J.}~\bibnamefont {Tang}},\ }\bibfield  {booktitle}
  {\emph {\bibinfo {booktitle} {Proceedings of the 25th ACM SIGKDD
  International Conference on Knowledge Discovery \& Data Mining}},\ }\href
  {\doibase 10.1145/3292500.3330982} {\ \bibinfo {series} {KDD '19},\ \bibinfo
  {pages} {723} (\bibinfo {year} {2019})}\BibitemShut {NoStop}%
\bibitem [{\citenamefont {Grattarola}\ \emph
  {et~al.}(2022{\natexlab{a}})\citenamefont {Grattarola}, \citenamefont
  {Zambon}, \citenamefont {Bianchi},\ and\ \citenamefont
  {Alippi}}]{Grattarola:2022wv}%
  \BibitemOpen
  \bibfield  {author} {\bibinfo {author} {\bibfnamefont {D.}~\bibnamefont
  {Grattarola}}, \bibinfo {author} {\bibfnamefont {D.}~\bibnamefont {Zambon}},
  \bibinfo {author} {\bibfnamefont {F.~M.}\ \bibnamefont {Bianchi}}, \ and\
  \bibinfo {author} {\bibfnamefont {C.}~\bibnamefont {Alippi}},\ }\bibfield
  {booktitle} {\emph {\bibinfo {booktitle} {IEEE Transactions on Neural
  Networks and Learning Systems}},\ }\href {\doibase
  10.1109/TNNLS.2022.3190922} {\bibfield  {journal} {\bibinfo  {journal} {IEEE
  Transactions on Neural Networks and Learning Systems}\ ,\ \bibinfo {pages}
  {1}} (\bibinfo {year} {2022}{\natexlab{a}})}\BibitemShut {NoStop}%
\bibitem [{\citenamefont {Wang}\ \emph {et~al.}(2020)\citenamefont {Wang},
  \citenamefont {Li}, \citenamefont {Ma}, \citenamefont {Montufar},
  \citenamefont {Zhuang},\ and\ \citenamefont {Fan}}]{wang2019haarpooling}%
  \BibitemOpen
  \bibfield  {author} {\bibinfo {author} {\bibfnamefont {Y.~G.}\ \bibnamefont
  {Wang}}, \bibinfo {author} {\bibfnamefont {M.}~\bibnamefont {Li}}, \bibinfo
  {author} {\bibfnamefont {Z.}~\bibnamefont {Ma}}, \bibinfo {author}
  {\bibfnamefont {G.}~\bibnamefont {Montufar}}, \bibinfo {author}
  {\bibfnamefont {X.}~\bibnamefont {Zhuang}}, \ and\ \bibinfo {author}
  {\bibfnamefont {Y.}~\bibnamefont {Fan}},\ }in\ \href
  {https://proceedings.mlr.press/v119/wang20m.html} {\emph {\bibinfo
  {booktitle} {Proceedings of the 37th International Conference on Machine
  Learning}}},\ \bibinfo {series} {Proceedings of Machine Learning Research},
  Vol.\ \bibinfo {volume} {119},\ \bibinfo {editor} {edited by\ \bibinfo
  {editor} {\bibfnamefont {H.~D.}\ \bibnamefont {III}}\ and\ \bibinfo {editor}
  {\bibfnamefont {A.}~\bibnamefont {Singh}}}\ (\bibinfo  {publisher} {PMLR},\
  \bibinfo {year} {2020})\ pp.\ \bibinfo {pages} {9952--9962}\BibitemShut
  {NoStop}%
\bibitem [{\citenamefont {Gilmer}\ \emph {et~al.}(2017)\citenamefont {Gilmer},
  \citenamefont {Schoenholz}, \citenamefont {Riley}, \citenamefont {Vinyals},\
  and\ \citenamefont {Dahl}}]{pmlr-v70-gilmer17a}%
  \BibitemOpen
  \bibfield  {author} {\bibinfo {author} {\bibfnamefont {J.}~\bibnamefont
  {Gilmer}}, \bibinfo {author} {\bibfnamefont {S.~S.}\ \bibnamefont
  {Schoenholz}}, \bibinfo {author} {\bibfnamefont {P.~F.}\ \bibnamefont
  {Riley}}, \bibinfo {author} {\bibfnamefont {O.}~\bibnamefont {Vinyals}}, \
  and\ \bibinfo {author} {\bibfnamefont {G.~E.}\ \bibnamefont {Dahl}},\ }in\
  \href {https://proceedings.mlr.press/v70/gilmer17a.html} {\emph {\bibinfo
  {booktitle} {Proceedings of the 34th International Conference on Machine
  Learning}}},\ \bibinfo {series} {Proceedings of Machine Learning Research},
  Vol.~\bibinfo {volume} {70},\ \bibinfo {editor} {edited by\ \bibinfo {editor}
  {\bibfnamefont {D.}~\bibnamefont {Precup}}\ and\ \bibinfo {editor}
  {\bibfnamefont {Y.~W.}\ \bibnamefont {Teh}}}\ (\bibinfo  {publisher} {PMLR},\
  \bibinfo {year} {2017})\ pp.\ \bibinfo {pages} {1263--1272}\BibitemShut
  {NoStop}%
\bibitem [{\citenamefont {Bacciu}\ \emph {et~al.}(2021)\citenamefont {Bacciu},
  \citenamefont {Conte}, \citenamefont {Grossi}, \citenamefont {Landolfi},\
  and\ \citenamefont {Marino}}]{Bacciu:2021wg}%
  \BibitemOpen
  \bibfield  {author} {\bibinfo {author} {\bibfnamefont {D.}~\bibnamefont
  {Bacciu}}, \bibinfo {author} {\bibfnamefont {A.}~\bibnamefont {Conte}},
  \bibinfo {author} {\bibfnamefont {R.}~\bibnamefont {Grossi}}, \bibinfo
  {author} {\bibfnamefont {F.}~\bibnamefont {Landolfi}}, \ and\ \bibinfo
  {author} {\bibfnamefont {A.}~\bibnamefont {Marino}},\ }\href {\doibase
  10.1007/s10618-021-00779-z} {\bibfield  {journal} {\bibinfo  {journal} {Data
  Mining and Knowledge Discovery}\ }\textbf {\bibinfo {volume} {35}},\ \bibinfo
  {pages} {2200} (\bibinfo {year} {2021})}\BibitemShut {NoStop}%
\bibitem [{\citenamefont {Auer}\ and\ \citenamefont
  {Bisseling}(2012)}]{auer2012graph}%
  \BibitemOpen
  \bibfield  {author} {\bibinfo {author} {\bibfnamefont {B.~F.}\ \bibnamefont
  {Auer}}\ and\ \bibinfo {author} {\bibfnamefont {R.~H.}\ \bibnamefont
  {Bisseling}},\ }\href@noop {} {\bibfield  {journal} {\bibinfo  {journal}
  {Graph Partitioning and Graph Clustering}\ }\textbf {\bibinfo {volume}
  {588}},\ \bibinfo {pages} {2} (\bibinfo {year} {2012})}\BibitemShut {NoStop}%
\bibitem [{\citenamefont {Pakhira}(2014)}]{pakhira2014linear}%
  \BibitemOpen
  \bibfield  {author} {\bibinfo {author} {\bibfnamefont {M.~K.}\ \bibnamefont
  {Pakhira}},\ }in\ \href@noop {} {\emph {\bibinfo {booktitle} {2014
  international conference on computational intelligence and communication
  networks}}}\ (\bibinfo {organization} {IEEE},\ \bibinfo {year} {2014})\ pp.\
  \bibinfo {pages} {1047--1051}\BibitemShut {NoStop}%
\bibitem [{\citenamefont {He}\ \emph {et~al.}(2016)\citenamefont {He},
  \citenamefont {Zhang}, \citenamefont {Ren},\ and\ \citenamefont
  {Sun}}]{he2016deep}%
  \BibitemOpen
  \bibfield  {author} {\bibinfo {author} {\bibfnamefont {K.}~\bibnamefont
  {He}}, \bibinfo {author} {\bibfnamefont {X.}~\bibnamefont {Zhang}}, \bibinfo
  {author} {\bibfnamefont {S.}~\bibnamefont {Ren}}, \ and\ \bibinfo {author}
  {\bibfnamefont {J.}~\bibnamefont {Sun}},\ }in\ \href@noop {} {\emph {\bibinfo
  {booktitle} {Proceedings of the IEEE conference on computer vision and
  pattern recognition}}}\ (\bibinfo {year} {2016})\ pp.\ \bibinfo {pages}
  {770--778}\BibitemShut {NoStop}%
\bibitem [{\citenamefont {Sj{\"o}strand}\ \emph {et~al.}(2006)\citenamefont
  {Sj{\"o}strand}, \citenamefont {Mrenna},\ and\ \citenamefont
  {Skands}}]{Sjostrand:2006vy}%
  \BibitemOpen
  \bibfield  {author} {\bibinfo {author} {\bibfnamefont {T.}~\bibnamefont
  {Sj{\"o}strand}}, \bibinfo {author} {\bibfnamefont {S.}~\bibnamefont
  {Mrenna}}, \ and\ \bibinfo {author} {\bibfnamefont {P.}~\bibnamefont
  {Skands}},\ }\bibfield  {booktitle} {\emph {\bibinfo {booktitle} {Journal of
  High Energy Physics}},\ }\href {\doibase 10.1088/1126-6708/2006/05/026} {\
  \textbf {\bibinfo {volume} {2006}},\ \bibinfo {pages} {026} (\bibinfo {year}
  {2006})}\BibitemShut {NoStop}%
\bibitem [{\citenamefont {Sj{\"o}strand}\ \emph {et~al.}(2015)\citenamefont
  {Sj{\"o}strand}, \citenamefont {Ask}, \citenamefont {Christiansen},
  \citenamefont {Corke}, \citenamefont {Desai}, \citenamefont {Ilten},
  \citenamefont {Mrenna}, \citenamefont {Prestel}, \citenamefont {Rasmussen},\
  and\ \citenamefont {Skands}}]{Sjostrand:2015vy}%
  \BibitemOpen
  \bibfield  {author} {\bibinfo {author} {\bibfnamefont {T.}~\bibnamefont
  {Sj{\"o}strand}}, \bibinfo {author} {\bibfnamefont {S.}~\bibnamefont {Ask}},
  \bibinfo {author} {\bibfnamefont {J.~R.}\ \bibnamefont {Christiansen}},
  \bibinfo {author} {\bibfnamefont {R.}~\bibnamefont {Corke}}, \bibinfo
  {author} {\bibfnamefont {N.}~\bibnamefont {Desai}}, \bibinfo {author}
  {\bibfnamefont {P.}~\bibnamefont {Ilten}}, \bibinfo {author} {\bibfnamefont
  {S.}~\bibnamefont {Mrenna}}, \bibinfo {author} {\bibfnamefont
  {S.}~\bibnamefont {Prestel}}, \bibinfo {author} {\bibfnamefont {C.~O.}\
  \bibnamefont {Rasmussen}}, \ and\ \bibinfo {author} {\bibfnamefont {P.~Z.}\
  \bibnamefont {Skands}},\ }\href {\doibase
  https://doi.org/10.1016/j.cpc.2015.01.024} {\bibfield  {journal} {\bibinfo
  {journal} {Computer Physics Communications}\ }\textbf {\bibinfo {volume}
  {191}},\ \bibinfo {pages} {159} (\bibinfo {year} {2015})}\BibitemShut
  {NoStop}%
\bibitem [{\citenamefont {Cacciari}\ \emph {et~al.}(2012)\citenamefont
  {Cacciari}, \citenamefont {Salam},\ and\ \citenamefont
  {Soyez}}]{Cacciari:2012us}%
  \BibitemOpen
  \bibfield  {author} {\bibinfo {author} {\bibfnamefont {M.}~\bibnamefont
  {Cacciari}}, \bibinfo {author} {\bibfnamefont {G.~P.}\ \bibnamefont {Salam}},
  \ and\ \bibinfo {author} {\bibfnamefont {G.}~\bibnamefont {Soyez}},\ }\href
  {\doibase 10.1140/epjc/s10052-012-1896-2} {\bibfield  {journal} {\bibinfo
  {journal} {The European Physical Journal C}\ }\textbf {\bibinfo {volume}
  {72}},\ \bibinfo {pages} {1896} (\bibinfo {year} {2012})}\BibitemShut
  {NoStop}%
\bibitem [{\citenamefont {Cacciari}\ \emph {et~al.}(2008)\citenamefont
  {Cacciari}, \citenamefont {Salam},\ and\ \citenamefont
  {Soyez}}]{Cacciari:2008wr}%
  \BibitemOpen
  \bibfield  {author} {\bibinfo {author} {\bibfnamefont {M.}~\bibnamefont
  {Cacciari}}, \bibinfo {author} {\bibfnamefont {G.~P.}\ \bibnamefont {Salam}},
  \ and\ \bibinfo {author} {\bibfnamefont {G.}~\bibnamefont {Soyez}},\
  }\bibfield  {booktitle} {\emph {\bibinfo {booktitle} {Journal of High Energy
  Physics}},\ }\href {\doibase 10.1088/1126-6708/2008/04/063} {\ \textbf
  {\bibinfo {volume} {2008}},\ \bibinfo {pages} {063} (\bibinfo {year}
  {2008})}\BibitemShut {NoStop}%
\bibitem [{\citenamefont {Kingma}\ and\ \citenamefont {Ba}(2014)}]{Adam}%
  \BibitemOpen
  \bibfield  {author} {\bibinfo {author} {\bibfnamefont {D.~P.}\ \bibnamefont
  {Kingma}}\ and\ \bibinfo {author} {\bibfnamefont {J.}~\bibnamefont {Ba}},\
  }\href {\doibase 10.48550/ARXIV.1412.6980} {\enquote {\bibinfo {title} {Adam:
  A method for stochastic optimization},}\ } (\bibinfo {year}
  {2014})\BibitemShut {NoStop}%
\bibitem [{\citenamefont {Goyal}\ \emph {et~al.}(2017)\citenamefont {Goyal},
  \citenamefont {Dollár}, \citenamefont {Girshick}, \citenamefont {Noordhuis},
  \citenamefont {Wesolowski}, \citenamefont {Kyrola}, \citenamefont {Tulloch},
  \citenamefont {Jia},\ and\ \citenamefont {He}}]{2017Accurate}%
  \BibitemOpen
  \bibfield  {author} {\bibinfo {author} {\bibfnamefont {P.}~\bibnamefont
  {Goyal}}, \bibinfo {author} {\bibfnamefont {P.}~\bibnamefont {Dollár}},
  \bibinfo {author} {\bibfnamefont {R.}~\bibnamefont {Girshick}}, \bibinfo
  {author} {\bibfnamefont {P.}~\bibnamefont {Noordhuis}}, \bibinfo {author}
  {\bibfnamefont {L.}~\bibnamefont {Wesolowski}}, \bibinfo {author}
  {\bibfnamefont {A.}~\bibnamefont {Kyrola}}, \bibinfo {author} {\bibfnamefont
  {A.}~\bibnamefont {Tulloch}}, \bibinfo {author} {\bibfnamefont
  {Y.}~\bibnamefont {Jia}}, \ and\ \bibinfo {author} {\bibfnamefont
  {K.}~\bibnamefont {He}},\ }\href@noop {} {\  (\bibinfo {year}
  {2017})}\BibitemShut {NoStop}%
\bibitem [{\citenamefont {Loshchilov}\ and\ \citenamefont
  {Hutter}(2016)}]{2016SGDR}%
  \BibitemOpen
  \bibfield  {author} {\bibinfo {author} {\bibfnamefont {I.}~\bibnamefont
  {Loshchilov}}\ and\ \bibinfo {author} {\bibfnamefont {F.}~\bibnamefont
  {Hutter}},\ }\href@noop {} {\  (\bibinfo {year} {2016})}\BibitemShut
  {NoStop}%
\bibitem [{\citenamefont {collaboration}\ \emph {et~al.}(2017)\citenamefont
  {collaboration} \emph {et~al.}}]{cms2017boosted}%
  \BibitemOpen
  \bibfield  {author} {\bibinfo {author} {\bibfnamefont {C.}~\bibnamefont
  {collaboration}} \emph {et~al.},\ }\href@noop {} {\bibfield  {journal}
  {\bibinfo  {journal} {Detector Performance Figures: CMS-DP-17-049}\ }
  (\bibinfo {year} {2017})}\BibitemShut {NoStop}%
\bibitem [{\citenamefont {Gallicchio}\ and\ \citenamefont
  {Schwartz}(2013)}]{Gallicchio:2013ww}%
  \BibitemOpen
  \bibfield  {author} {\bibinfo {author} {\bibfnamefont {J.}~\bibnamefont
  {Gallicchio}}\ and\ \bibinfo {author} {\bibfnamefont {M.~D.}\ \bibnamefont
  {Schwartz}},\ }\href {\doibase 10.1007/JHEP04(2013)090} {\bibfield  {journal}
  {\bibinfo  {journal} {Journal of High Energy Physics}\ }\textbf {\bibinfo
  {volume} {2013}},\ \bibinfo {pages} {90} (\bibinfo {year}
  {2013})}\BibitemShut {NoStop}%
\bibitem [{\citenamefont {Grattarola}\ \emph
  {et~al.}(2022{\natexlab{b}})\citenamefont {Grattarola}, \citenamefont
  {Zambon}, \citenamefont {Bianchi},\ and\ \citenamefont
  {Alippi}}]{Grattarola2022}%
  \BibitemOpen
  \bibfield  {author} {\bibinfo {author} {\bibfnamefont {D.}~\bibnamefont
  {Grattarola}}, \bibinfo {author} {\bibfnamefont {D.}~\bibnamefont {Zambon}},
  \bibinfo {author} {\bibfnamefont {F.~M.}\ \bibnamefont {Bianchi}}, \ and\
  \bibinfo {author} {\bibfnamefont {C.}~\bibnamefont {Alippi}},\ }\href
  {\doibase 10.1109/TNNLS.2022.3190922} {\bibfield  {journal} {\bibinfo
  {journal} {IEEE Transactions on Neural Networks and Learning Systems}\ ,\
  \bibinfo {pages} {1}} (\bibinfo {year} {2022}{\natexlab{b}})}\BibitemShut
  {NoStop}%
\bibitem [{\citenamefont {Mikuni}\ and\ \citenamefont
  {Canelli}(2020)}]{mikuni2020abcnet}%
  \BibitemOpen
  \bibfield  {author} {\bibinfo {author} {\bibfnamefont {V.}~\bibnamefont
  {Mikuni}}\ and\ \bibinfo {author} {\bibfnamefont {F.}~\bibnamefont
  {Canelli}},\ }\href@noop {} {\bibfield  {journal} {\bibinfo  {journal} {The
  European Physical Journal Plus}\ }\textbf {\bibinfo {volume} {135}},\
  \bibinfo {pages} {1} (\bibinfo {year} {2020})}\BibitemShut {NoStop}%
\end{thebibliography}%

%\clearpage

\end{document}